\documentclass[aps,prl,twocolumn,superscriptaddress, 10pt]{revtex4-1}
\usepackage{graphics,graphicx, array, afterpage}
\usepackage{qcircuit,amsmath}
\usepackage{grffile}
\usepackage[export]{adjustbox}
\usepackage{longtable}

\newcommand{\ConceptFigure}{
\begin{figure*}[t]
\centering
\begin{tabular}[c]{l l  l}
\multicolumn{3}{l}{\bf{(a) Grover Search Algorithm}} \\
&&\\

\multicolumn{3}{c}{\includegraphics[width=\textwidth]{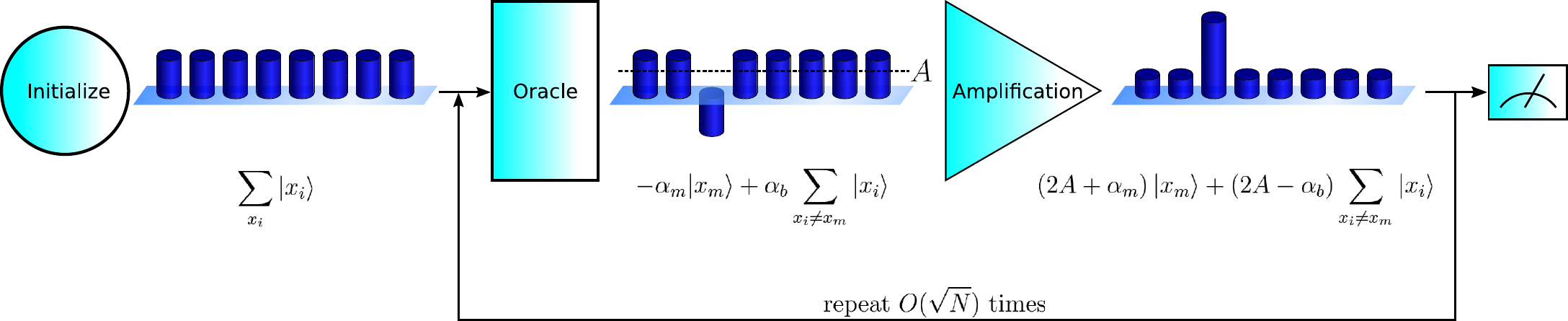}}
\\
&&\\

\bf{(b) Boolean Oracle} && \bf{(c) Example Single-Solution Boolean Oracle} \\

\Qcircuit @C=0.7em @R=0.2em @!R {
& &  	&						\mbox{Init}	&					&&\mbox{Amplification}&&&&\\		
\push{|q_1\rangle:} &|0\rangle & & \gate{H}	&\multigate{3}{\text{Oracle}} 	&\gate{H}		& \gate{X}	& \ctrl{2}	& \gate{X}	 	&\gate{H}& \qw 	 \\
\push{|q_2\rangle:} & |0\rangle & & \gate{H}	&\ghost{\text{Oracle}}		& \gate{H}		& \gate{X}	& \ctrl{1}	& \gate{X}		&\gate{H} & \qw 	 \\
\push{|q_3\rangle:} & |0\rangle & & \gate{H}	&\ghost{\text{Oracle}}		& \gate{H}		& \gate{X}	& \gate{Z}	& \gate{X}		&\gate{H}& \qw 	 \\
\push{|q_a\rangle:} & |1\rangle & 	&\gate{H}&\ghost{\text{Oracle}}		& \gate{H}	 	& \qw		& \qw	& \qw  		&\qw & \qw  
\\
}

&& \enspace

\Qcircuit @C=0.7em @R=0.2em @!R {
& & \mbox{Init} 		&&\mbox{Oracle}  &			&&\mbox{Amplification}&&&&\\		
|0\rangle & & \gate{H}	&\gate{X}	&\ctrl{3} 	&\gate{X}	&\gate{H}	& \gate{X}	& \ctrl{2}	& \gate{X}	&\gate{H}& \qw \gategroup{2}{12}{4}{13}{0.5em}{\}}& &	 \\
|0\rangle & & \gate{H}	&\qw		&\ctrl{2}	&\qw		& \gate{H}	& \gate{X}	& \ctrl{1}	& \gate{X}	&\gate{H} & \qw & & \push{\frac{5}{4\sqrt{2}} |011\rangle + \frac{1}{4\sqrt{2}}\sum_{x \neq 011}|x\rangle}	 \\
|0\rangle & & \gate{H}	&\qw		&\ctrl{1}	&\qw		& \gate{H}	& \gate{X}	& \gate{Z}	& \gate{X}	&\gate{H}& \qw & &  \\
|1\rangle & & \gate{H}	&\qw		&\targ		&\qw		& \gate{H}	& \qw		& \qw		& \qw  	&\qw & \qw  &|1\rangle
\gategroup{2}{4}{5}{6}{.7em}{--} & 
}

\\
\\

\bf{(d) Phase Oracle} & & \bf{(e) Example Two-Solution Phase Oracle}   \\

\Qcircuit @C=0.7em @R=0.2em @!R {
& &  						\mbox{Init}						&& &\mbox{Amplification}&&&&&&\\		
\push{|q_1\rangle:} & |0\rangle & & \gate{H}	&\multigate{2}{\text{Oracle}} 	&\gate{H}		& \gate{X}	& \ctrl{2}	& \gate{X}	 	&\gate{H}& \qw 	 \\
\push{|q_2\rangle:} & |0\rangle & & \gate{H}	&\ghost{\text{Oracle}}		& \gate{H}		& \gate{X}		& \ctrl{1}	& \gate{X}		&\gate{H} & \qw 	 \\
\push{|q_3\rangle:} &  |0\rangle & & \gate{H}	&\ghost{\text{Oracle}}		& \gate{H}		& \gate{X}		& \gate{Z}	& \gate{X}		&\gate{H}& \qw  \\}

&& \enspace

\Qcircuit @C=0.7em @R=0.2em @!R {
& & \mbox{Init} 		&\mbox{\hspace{1.2em} Oracle} &  			&&\mbox{Amplification}&&&&&&\\		
|0\rangle & & \gate{H}	&\ctrl{2}	&\qw 		&\gate{H}	& \gate{X}	& \ctrl{2}	& \gate{X}	&\gate{H}& \qw \gategroup{2}{11}{4}{12}{0.5em}{\}}& &	 \\
|0\rangle & & \gate{H}	&\qw		&\ctrl{1}	& \gate{H}	& \gate{X}	& \ctrl{1}	& \gate{X}	&\gate{H} & \qw & & \push{\frac{1}{\sqrt{2}} \left(|011\rangle+|101\rangle\right)}	 \\
|0\rangle & & \gate{H}	&\gate{Z}	&\gate{Z}	& \gate{H}	& \gate{X}	& \gate{Z}	& \gate{X}	&\gate{H} & \qw \gategroup{2}{4}{4}{5}{0.7em}{--}  & &  
}
\\
\\
\\

\end{tabular}
\caption{(a) Evolution of relative amplitudes for each state during a Grover search algorithm. The initialization stage creates an equal superposition of all possible input states, so the amplitude $\alpha_i=1$ for all basis states $|x_i\rangle$. The oracle stage marks the desired state, so the amplitude $\alpha_m$ of the marked state $|x_m\rangle$ becomes negative while the amplitudes $\alpha_b$ of the unmarked states $|x_i\rangle, x_i \neq x_m$ remain unchanged. The amplification stage performs a reflection about the mean vector $\sum_{x_i} |x_i\rangle$, which has amplitude $A = \frac{1}{N} \sum_i \alpha_i$, to amplify the marked state. An appropriate number of repetitions of the oracle and amplification stages will maximize the amplitude of the correct answer. All qubit states are normalized by the factor $\frac{1}{\sqrt{N}}$. The algorithm can also be generalized to mark and amplify the amplitude of $t$ desired states. 
(b) General circuit diagram for a Grover search algorithm using a Boolean oracle, depicted using standard quantum circuit diagram notation \cite{NielsenChuang11}.
The last qubit $q_a$ is the ancilla qubit. 
(c) Example single-solution Boolean oracle marking the $|011\rangle$ state. 
(d) General circuit diagram for a Grover search algorithm using a phase oracle.
(e) Example two-solution phase oracle marking the $|011\rangle$ and $|101\rangle$ states.}
\label{fig:GroverConcept}
\end{figure*}
}

\newcommand{\ToffCompressedCircuit}{
\begin{tabular}[c]{l l}

\Qcircuit @C=.4em @R=1.15em @!R {
\lstick{|q_1\rangle} & \ctrl{1} 	& \qw &\\
\lstick{|q_2\rangle} & \ctrl{1} 	& \qw &\push{=}\\
\lstick{|q_t\rangle} & \targ  &\qw &
}
& 
\Qcircuit @C=.4em @R=0.3em @!R {
   & \gate{R_y} & \gate{R_x}	& \gate{XX(\frac{\pi}{8})}    & \qw &\qw & \multigate{1}{XX(\frac{\pi}{4})} & \qw \\
 & \gate{R_y} &  \gate{R_x} &  \qw  \qwx \qwx[1] & \multigate{1}{XX(\frac{\pi}{8})}  & \gate{R_{\phi}} & \ghost{XX(\frac{\pi}{4})} & \qw \\
 &  \gate{R_x}  &\qw  & \gate{XX(\frac{\pi}{8})}	 & \ghost{XX(\frac{\pi}{8})}  &\qw  & \qw & \qw 	
}
\\ 
& \\

\Qcircuit @C=.4em @R=1.8em @!R {
\lstick{|q_1\rangle}\\
\lstick{|q_2\rangle}\\
\lstick{|q_t\rangle}
}
&
\Qcircuit @C=.4em @R=0.3em @!R {
\lstick{\cdots}  & \qw & \qw  & \qw & \multigate{1}{XX(\frac{\pi}{4})} & \gate{R_y} & \qw \\
\lstick{\cdots}  & \gate{R_{\phi}} & \multigate{1}{XX(\frac{\pi}{8})} & \gate{R_{\phi}} & \ghost{XX(\frac{\pi}{4})} & \qw & \qw \\
 \lstick{\cdots}  & \qw & \ghost{XX(\frac{\pi}{8})} & \qw  & \qw  & \qw  & \qw
}

\end{tabular}
}

\newcommand{\ToffTestCircuit}{

\Qcircuit @C=1em @R=1em @!R {
\\
\lstick{|q_1\rangle}  & \gate{ R_y \left( \pm\frac{\pi}{2} \right)}  & \ctrl{1} 	& \gate{ R_y \left(\pm \frac{\pi}{2} \right)} & \qw &\\
\lstick{|q_2\rangle} & \gate{ R_y \left( \pm\frac{\pi}{2} \right)}   & \ctrl{1} 	& \gate{ R_y \left( \pm\frac{\pi}{2} \right)} & \qw &\\
\lstick{|q_t\rangle} & \gate{ R_y \left( \pm\frac{\pi}{2} \right)}    & \targ  	& \gate{ R_y \left( \pm\frac{\pi}{2} \right)} &\qw &
}

}

\newcommand{\ZCircuit}{
\begin{figure}[h]
\centering
\begin{tabular}[c]{l  l}

\Qcircuit @C=.5em @R=.7em @!R {
\lstick{|q\rangle} & \gate{R_z(\theta)} 	& \qw &\push{\rule{.3em}{0em}=} 
}
&
\Qcircuit @C=.5em @R=0.3em @!R {
   & \gate{R_y(\frac{\pi}{2})} & \gate{R_x(\theta)}	 & \gate{R_y(-\frac{\pi}{2})} & \qw \\
}
\\

\end{tabular}
\caption{$R_z(\theta)$ gate implementation using $R_x(\theta)$ and $R_y(\theta)$ gates.} 
\label{fig:ZCircuit}
\end{figure}
}
\newcommand{\TwoQubitCircuits}{
\begin{figure}
\centering
\begin{tabular}[c]{c}
\multicolumn{1}{l}{\bf{(a) Controlled-$NOT$ Gate}}\\
\\

\begin{tabular}[c]{l c  l}

\Qcircuit @C=.5em @R=1.15em @!R {
\lstick{|q_c\rangle} & \ctrl{1} 	& \qw \\
\lstick{|q_t\rangle} & \targ 	& \qw
}
&\raisebox{-3.5\totalheight}{=}&
\Qcircuit @C=.5em @R=0.3em @!R {
& \gate{R_y(\alpha\frac{\pi}{2})} &\multigate{1}{XX(\alpha\frac{\pi}{4})} 	&\gate{R_y(-\alpha\frac{\pi}{2})} &\gate{R_z(-\frac{\pi}{2})} & \qw \\
& \qw 					  &\ghost{XX(\alpha\frac{\pi}{4})} 		& \gate{R_x(-\frac{\pi}{2})} & \qw 	& \qw 
}
\\ 

\end{tabular}
\\
\\
\multicolumn{1}{l}{\bf{(b) Controlled-$Z$ Gate}}\\
\\
\begin{tabular}[c]{l c  l}

\Qcircuit @C=.5em @R=.7em @!R {
\lstick{|q_c\rangle} & \ctrl{1} 	& \qw \\
\lstick{|q_t\rangle} & \gate{Z} 	& \qw
}
&\raisebox{-3.5\totalheight}{=}&
\Qcircuit @C=.5em @R=0.3em @!R {
   & \gate{R_y(-\frac{\pi}{2})} & \gate{R_x(-\frac{\pi}{2})}	 & \multigate{1}{XX(\alpha\frac{\pi}{4})} & \gate{R_y(\frac{\pi}{2})} & \qw \\
 & \gate{R_y(\alpha\frac{\pi}{2})} &  \gate{R_x(\alpha\frac{\pi}{2})}  & \ghost{XX(\alpha\frac{\pi}{4})} &  \gate{R_y(-\alpha\frac{\pi}{2})} & \qw 	
}
\\ 

\end{tabular}
\end{tabular}
\caption{Two-qubit composite gates. $\chi_{ct}$ is the parameter for the $XX$ gate between the two qubits. Let $\alpha = \text{sgn}(\chi_{ct})$. (a) $CNOT$ gate implementation using $XX(\chi)$, $R_x(\theta)$, $R_y(\theta)$, and $R_z(\theta)$ gates.  (b) Controlled-$Z$ gate implementation using $XX(\chi)$, $R_x(\theta)$, and $R_y(\theta)$ gates.}
\label{fig:2QCircuits}
\end{figure}
}
\newcommand{\ThreeQubitCircuits}{
\begin{figure*}
\centering
\begin{tabular}[c]{c}
\multicolumn{1}{l}{\bf{(a) Toffoli-3 Gate}}\\
\\

\begin{tabular}[c]{l l}

\Qcircuit @C=.5em @R=1.3em @!R {
\lstick{|q_1\rangle} & \ctrl{1} 	& \qw &\\
\lstick{|q_2\rangle} & \ctrl{1} 	& \qw &\push{\rule{.3em}{0em}=}  \\
\lstick{|q_t\rangle} & \targ  &\qw &
}
& 
\Qcircuit @C=.5em @R=0.3em @!R {
   & \gate{R_y(-\beta\frac{\pi}{2})} & \gate{R_x(\beta\frac{3\pi}{4})}	& \gate{XX(\beta\frac{\pi}{8})}    & \qw &\qw & \multigate{1}{XX(\alpha\frac{\pi}{4})} & \qw \\
 & \gate{R_y(-\gamma\frac{\pi}{2})} &  \gate{R_x(\gamma\frac{\pi}{2})} &  \qw  \qwx \qwx[1] & \multigate{1}{XX(\gamma\frac{\pi}{8})}  & \gate{R(-\frac{2\pi}{3}, (\frac{\gamma+1}{2})\pi - P)} & \ghost{XX(\alpha\frac{\pi}{4})} & \qw \\
 &  \gate{R_x(\frac{\pi}{4})}  &\qw  & \gate{XX(\beta\frac{\pi}{8})}	 & \ghost{XX(\gamma\frac{\pi}{8})}  &\qw  & \qw & \qw 	
}
\\ 
& \\

\Qcircuit @C=.5em @R=2.0em @!R {
\lstick{|q_1\rangle}\\
\lstick{|q_2\rangle}\\
\lstick{|q_t\rangle}
}
&
\Qcircuit @C=.5em @R=0.3em @!R {
\lstick{\cdots}  & \qw & \qw  & \qw & \multigate{1}{XX(\alpha\frac{\pi}{4})} & \gate{R_y(\beta\frac{\pi}{2})} & \qw \\
\lstick{\cdots}  & \gate{R(-\alpha\beta\gamma\frac{2\pi}{3}, (\frac{\alpha\beta+1}{2})\pi - \alpha\beta\gamma P)} & \multigate{1}{XX(\gamma\frac{\pi}{8})} & \gate{R(\pi,-\alpha\beta\gamma\frac{\pi}{4})} & \ghost{XX(\alpha\frac{\pi}{4})} & \qw & \qw \\
 \lstick{\cdots}  & \qw & \ghost{XX(\gamma\frac{\pi}{8})} & \qw  & \qw  & \qw  & \qw
}

\end{tabular}
\\
\\
\multicolumn{1}{l}{\bf{(b) $CCZ$ Gate}}\\
\\
\begin{tabular}[c]{l l}

\Qcircuit @C=.5em @R=.85em @!R {
\lstick{|q_1\rangle} & \ctrl{1} 	& \qw &\\
\lstick{|q_2\rangle} & \ctrl{1} 	& \qw &\push{\rule{.3em}{0em}=}  \\
\lstick{|q_t\rangle} & \gate{Z}  &\qw &
}
& 
\Qcircuit @C=.5em @R=0.3em @!R {
   & \gate{R_y(-\beta\frac{\pi}{2})} & \gate{R_x(\beta\frac{3\pi}{4})}	& \gate{XX(\beta\frac{\pi}{8})}    & \qw &\qw & \multigate{1}{XX(\alpha\frac{\pi}{4})} & \qw \\
 & \gate{R_y(-\gamma\frac{\pi}{2})} &  \gate{R_x(\gamma\frac{\pi}{2})} &  \qw  \qwx \qwx[1] & \multigate{1}{XX(\gamma\frac{\pi}{8})}  & \gate{R(-\frac{2\pi}{3}, (\frac{\gamma+1}{2})\pi - P)} & \ghost{XX(\alpha\frac{\pi}{4})} & \qw \\
&\gate{R_y(\frac{\pi}{2})} &  \gate{R_x(\frac{\pi}{4})}   & \gate{XX(\beta\frac{\pi}{8})}	 & \ghost{XX(\gamma\frac{\pi}{8})}  &\qw  & \qw & \qw 	
}
\\ 
& \\

\Qcircuit @C=.5em @R=2.0em @!R {
\lstick{|q_1\rangle}\\
\lstick{|q_2\rangle}\\
\lstick{|q_t\rangle}
}
&
\Qcircuit @C=.5em @R=0.3em @!R {
\lstick{\cdots}  & \qw & \qw  & \qw & \multigate{1}{XX(\alpha\frac{\pi}{4})} & \gate{R_y(\beta\frac{\pi}{2})} & \qw \\
\lstick{\cdots}  & \gate{R(-\alpha\beta\gamma\frac{2\pi}{3}, (\frac{\alpha\beta+1}{2})\pi - \alpha\beta\gamma P)} & \multigate{1}{XX(\gamma\frac{\pi}{8})} & \gate{R(\pi,-\alpha\beta\gamma\frac{\pi}{4})} & \ghost{XX(\alpha\frac{\pi}{4})} & \qw & \qw \\
 \lstick{\cdots}  & \qw & \ghost{XX(\gamma\frac{\pi}{8})} & \qw  & \qw  & \gate{R_y(-\frac{\pi}{2})}  & \qw
}

\end{tabular}

\end{tabular}
\caption{Three-qubit composite gates using $XX(\chi)$, $R_x(\theta)$, $R_y(\theta)$, and $R(\theta,\phi)$ gates. Let $\alpha =\text{sgn}(\chi_{12})$, $\beta = \text{sgn}(\chi_{1t})$, $\gamma = \text{sgn}(\chi_{2t})$, and $P = \arcsin{ \left(\sqrt{\frac{2}{3}} \right)}$. (a) Toffoli-3 gate implementation. (b) Controlled-controlled-$Z$ ($CCZ$) gate implementation.} 
\label{fig:3QCircuits}
\end{figure*}
}

\newcommand{\ToffFourCircuit}{
\begin{figure*}
\centering
\begin{tabular}[c]{l l}

\Qcircuit @C=.5em @R=1.18em @!R {
\lstick{|q_1\rangle} & \ctrl{3} 	& \qw &\\
\lstick{|q_2\rangle} & \ctrl{2} 	& \qw &  \\
\lstick{|q_3\rangle} & \ctrl{1}  &\qw &\push{\rule{.3em}{0em}=}\\
\lstick{|q_t\rangle} & \targ  &\qw &\\
\lstick{|q_a\rangle} & \qw  &\qw &\\
}
& 
\Qcircuit @C=.5em @R=0.3em @!R {
&\gate{R_y(\alpha_1\alpha_2\frac{\pi}{2})} 	& \qw						& \qw					&\gate{XX(\alpha_1\frac{\pi}{4})}  		&\qw 						&\qw 						&\qw 						 & \qw \\
&\gate{R_y(\frac{\pi}{2})} 			&\gate{XX(\alpha_2\frac{\pi}{4})}  	& \qw					&\qw \qwx \qwx[1] 				&\qw  						& \qw 						& \gate{XX(\alpha_2\frac{\pi}{4})}				 & \qw \\
& \qw 							&\qw\qwx \qwx[1]  			& \qw					&\qw \qwx \qwx[1]	 				&\qw  						&\qw  						& \qw\qwx \qwx[1] 			& \qw 	\\
&\qw							&\qw\qwx \qwx[1]				& \qw					&\qw \qwx \qwx[1]					&\qw 						&\qw 						&\qw\qwx \qwx[1]				&\qw\\
&\gate{R_y(\alpha_t\frac{\pi}{4})}  		&\gate{XX(\alpha_2\frac{\pi}{4})}  	& \gate{R(-\alpha_t \pi,Q\pi)} 	&\gate{XX(\alpha_1\frac{\pi}{4})}	&\gate{R_y(\alpha_t\frac{\pi}{4})}  	&\gate{R_x(-\alpha_2\frac{\pi}{2})} 	 & \gate{XX(\alpha_2\frac{\pi}{4})}	 & \qw
}
\\ 
& \\

\Qcircuit @C=.5em @R=2.0em @!R {
\lstick{|q_1\rangle}\\
\lstick{|q_2\rangle}\\
\lstick{|q_3\rangle}\\
\lstick{|q_t\rangle}\\
\lstick{|q_a\rangle}\\
}
& 
\Qcircuit @C=.5em @R=0.3em @!R {
\lstick{\cdots} &\qw 					& \qw						& \qw											&\qw  						&\qw 									&\qw 						&\qw  \\
\lstick{\cdots} &\qw 					&\qw  						& \qw											&\qw 						&\qw  									& \qw 						 & \qw \\
\lstick{\cdots} & \gate{R_y(-\beta\frac{\pi}{2})} 	&\gate{R_x(\beta\frac{\pi}{2})}  	& \qw									&\multigate{1}{XX(\beta\frac{\pi}{8})} 	&\gate{R(-\frac{2\pi}{3}, (\frac{\beta+1}{2})\pi - P)} 	&\gate{XX(\alpha_3\frac{\pi}{4})}  	& \qw \\
\lstick{\cdots} &\gate{R_x(\frac{\pi}{4})}		&\qw						& \multigate{1}{XX(\alpha_t\frac{\pi}{8})} 	&\ghost{XX(\beta\frac{\pi}{8})}		&\qw 									&\qw \qwx \qwx[1] 			&\qw\\
\lstick{\cdots} &\gate{R_y(-\alpha_t\frac{\pi}{4})} 	&\gate{R_x(\alpha_t\frac{3\pi}{4})}  	& \ghost{XX(\alpha_t\frac{\pi}{8})} 						&\qw						&\qw  									&\gate{XX(\alpha_3\frac{\pi}{4})} 	 & \qw
}
\\ 
& \\

\Qcircuit @C=.5em @R=2.0em @!R {
\lstick{|q_1\rangle}\\
\lstick{|q_2\rangle}\\
\lstick{|q_3\rangle}\\
\lstick{|q_t\rangle}\\
\lstick{|q_a\rangle}\\
}
&
\Qcircuit @C=.5em @R=0.3em @!R {
\lstick{\cdots} &\qw 																& \qw						& \qw										&\qw  						&\gate{R_x(-\alpha_1\pi)} 	&\qw  \\
\lstick{\cdots} &\qw 																&\qw  						& \qw										&\qw 						&\qw  					& \qw \\
\lstick{\cdots} & \gate{R(-\alpha_3\alpha_t\beta\frac{2\pi}{3}, (\frac{\alpha_3\alpha_t+1}{2})\pi - \alpha_3\alpha_t\beta P)} &\multigate{1}{XX(\beta\frac{\pi}{8})}  & \gate{R(\pi,-\alpha_3\alpha_t\beta\frac{\pi}{4})}		&\gate{XX(\alpha_3\frac{\pi}{4})} 	&\qw 					& \qw \\
\lstick{\cdots} &\qw																	&\ghost{XX(\beta\frac{\pi}{8})}		& \qw 										&\qw \qwx \qwx[1]				&\qw 					&\qw\\
\lstick{\cdots} &\qw 																&\qw  						& \qw 										&\gate{XX(\alpha_3\frac{\pi}{4})}	&\gate{R_y(\alpha_t\frac{\pi}{4})}  & \qw
}

\\ 
& \\

\Qcircuit @C=.5em @R=2.0em @!R {
\lstick{|q_1\rangle}\\
\lstick{|q_2\rangle}\\
\lstick{|q_3\rangle}\\
\lstick{|q_t\rangle}\\
\lstick{|q_a\rangle}\\
}
&
\Qcircuit @C=.5em @R=0.3em @!R {
\lstick{\cdots} & \qw					& \qw					&\gate{XX(\alpha_1\frac{\pi}{4})}  	&\qw 						&\qw 						&\qw 						& \gate{R_y(-\alpha_1\alpha_2\frac{\pi}{2})} & \qw \\
\lstick{\cdots} &\gate{XX(\alpha_2\frac{\pi}{4})}  	& \qw					&\qw \qwx \qwx[1] 			&\qw  						& \qw 						&\gate{XX(\alpha_2\frac{\pi}{4})}	& \gate{R_y(-\frac{\pi}{2})}			 & \qw \\
\lstick{\cdots} &\qw\qwx \qwx[1]  			& \qw					&\qw \qwx \qwx[1]	 			&\qw  						&\qw  						& \qw\qwx \qwx[1] 			& \qw 							& \qw 	\\
\lstick{\cdots} &\qw\qwx \qwx[1]			& \qw					&\qw \qwx \qwx[1]				&\qw 						&\qw 						&\qw\qwx \qwx[1]				& \qw 							&\qw\\
\lstick{\cdots} &\gate{XX(\alpha_2\frac{\pi}{4})}  	& \gate{R(-\alpha_t \pi,Q\pi)} 	&\gate{XX(\alpha_1\frac{\pi}{4})}	&\gate{R_y(\alpha_t\frac{\pi}{4})}  	&\gate{R_x(-\alpha_2\frac{\pi}{2})} 	&\gate{XX(\alpha_2\frac{\pi}{4})}	& \gate{R_y(\alpha_t\frac{\pi}{4})} 		& \qw
}

\end{tabular}
\caption{Toffoli-4 gate implementation using $XX(\chi)$, $R_x(\theta)$, $R_y(\theta)$, and $R(\theta,\phi)$ gates. Let $\alpha_1 = \text{sgn}(\chi_{1a})$, $\alpha_2 = \text{sgn}(\chi_{2a})$, $\alpha_3 = \text{sgn}(\chi_{3a})$, $\alpha_t = \text{sgn}(\chi_{ta})$, $\beta = \text{sgn}(\chi_{3t})$, $P = \arcsin{ \left(\sqrt{\frac{2}{3}} \right)}$, and $Q = \frac{1}{8}(4-3\alpha_2\alpha_t)$.} 
\label{fig:Toff4Circuit}
\end{figure*}
}

\newcommand{\GroverCircuits}{
\begin{figure*}
\centering
\begin{tabular}[c]{c}
\multicolumn{1}{l}{\bf{(a) Grover Initialization Stage Implementation}}\\
\\
\begin{tabular}[c]{l l}

\Qcircuit @C=.5em @R=2.0em @!R {
\lstick{|q_1\rangle:}\\
\lstick{|q_2\rangle:}\\
\lstick{|q_3\rangle:}\\
\lstick{|q_a\rangle:}
}
& 
\Qcircuit @C=.5em @R=0.3em @!R {
&&\lstick{|0\rangle}  &\gate{R_x(\pi)} 	&\gate{R_y(-\frac{\pi}{2})}	& \qw \\
&&\lstick{|0\rangle}  &\gate{R_x(\pi)} 	&\gate{R_y(-\frac{\pi}{2})}	& \qw \\
&&\lstick{|0\rangle}  &\gate{R_x(\pi)} 	&\gate{R_y(-\frac{\pi}{2})}	& \qw \\
&&\lstick{|0\rangle}  &\qw		 &\gate{R_y(-\frac{\pi}{2})}	&\qw	
}
\\ 

\end{tabular}
\\
\\
\multicolumn{1}{l}{\bf{(b) Grover Amplification Stage Implementation}}\\
\\

\begin{tabular}[c]{l l}

\Qcircuit @C=.5em @R=2.0em @!R {
\lstick{|q_1\rangle}\\
\lstick{|q_2\rangle}\\
\lstick{|q_3\rangle}\\
\lstick{|q_a\rangle}
}
& 
\Qcircuit @C=.5em @R=0.3em @!R {
   & \gate{R_y\left((1-\beta)\frac{\pi}{2}\right)} & \gate{R_x(\beta\frac{3\pi}{4})}	& \gate{XX(\beta\frac{\pi}{8})}    & \qw &\qw & \multigate{1}{XX(\alpha\frac{\pi}{4})} & \qw \\
 & \gate{R_y\left((1-\gamma)\frac{\pi}{2}\right)} &  \gate{R_x(\gamma\frac{\pi}{2})} &  \qw  \qwx \qwx[1] & \multigate{1}{XX(\gamma\frac{\pi}{8})}  & \gate{R(-\frac{2\pi}{3}, (\frac{\gamma+1}{2})\pi - P)} & \ghost{XX(\alpha\frac{\pi}{4})} & \qw \\
&\gate{R_y(\pi)} &  \gate{R_x(\frac{\pi}{4})}   & \gate{XX(\beta\frac{\pi}{8})}	 & \ghost{XX(\gamma\frac{\pi}{8})}  &\qw  & \qw & \qw \\
&\gate{R_y(\frac{\pi}{2})}	&\gate{R_x(\pi)}	&\qw	&\qw	&\qw	&\qw	&\qw	
}
\\ 
& \\

&
\Qcircuit @C=.5em @R=0.3em @!R {
\lstick{\cdots}  & \qw & \qw  & \qw & \multigate{1}{XX(\alpha\frac{\pi}{4})} & \gate{R_y\left((\beta-1)\frac{\pi}{2}\right)} & \qw \\
\lstick{\cdots}  & \gate{R(-\alpha\beta\gamma\frac{2\pi}{3}, (\frac{\alpha\beta+1}{2})\pi - \alpha\beta\gamma P)} & \multigate{1}{XX(\gamma\frac{\pi}{8})} & \gate{R(\pi,-\alpha\beta\gamma\frac{\pi}{4})} & \ghost{XX(\alpha\frac{\pi}{4})} & \gate{R_y(-\frac{\pi}{2})} & \qw \\
 \lstick{\cdots}  & \qw & \ghost{XX(\gamma\frac{\pi}{8})} & \qw  & \qw  & \gate{R_y(\pi)}  & \qw \\
 \lstick{\cdots} &\qw	&\qw	&\qw	&\qw	&\qw	&\qw
}

\end{tabular}
\end{tabular}
\caption{Grover search algorithm implementation by substage using $XX(\chi)$, $R_x(\theta)$, $R_y(\theta)$, and $R(\theta,\phi)$ gates. The circuits shown are for use with Boolean oracles; removing the ancilla qubit $|q_a\rangle$ produces the necessary circuits for use with a phase oracle. Let $\alpha = \text{sgn}(\chi_{12})$, $\beta = \text{sgn}(\chi_{1t})$, $\gamma = \text{sgn}(\chi_{2t})$, and $P = \arcsin{  \left(\sqrt{\frac{2}{3}} \right) }$. (a) Grover initialization stage implementation. (b) Grover amplification stage implementation.} 
\label{fig:GroverCircuits}
\end{figure*}
}

\newcommand{\ChooseOneOracles}{

\begin{center}
\centering
\begin{longtable*}[c]{| p{1.1cm} | c c | c c | p{1.1cm} | c c | c c |}
\hline
Marked && Boolean Oracle && Phase Oracle & Marked && Boolean Oracle && Phase Oracle\\
\hline\hline
\endhead
\bf{000} 
&&\enspace

\Qcircuit @C=0.5em @R=0.2em @!R {
|q_1\rangle &&& \gate{X}		&\ctrl{3}	& \gate{X}	& \qw 	 \\
|q_2\rangle &&& \gate{X}		&\ctrl{2}	& \gate{X}	& \qw 	 \\
|q_3\rangle &&& \gate{X}		&\ctrl{1}	& \gate{X}	& \qw 	 \\
|q_a\rangle &&& \qw 		&\targ		& \qw 		& \qw  \\ \\
}

&&\enspace

\Qcircuit @C=0.5em @R=0.2em @!R {
|q_1\rangle &&& \gate{X}	&\ctrl{2}	&\gate{X}	& \qw \\
|q_2\rangle &&& \gate{X}	&\ctrl{1}	&\gate{X}	& \qw   \\
|q_3\rangle &&& \gate{X}	&\gate{Z}	&\gate{X}	& \qw \\ \\
}

&
\bf{100} 
&&\enspace

\Qcircuit @C=0.5em @R=0.2em @!R {
|q_1\rangle &&& \qw		&\ctrl{3}	& \qw		& \qw 	 \\
|q_2\rangle &&& \gate{X}		&\ctrl{2}	& \gate{X}	& \qw 	 \\
|q_3\rangle &&& \gate{X}		&\ctrl{1}	& \gate{X}	& \qw 	 \\
|q_a\rangle &&& \qw	 	&\targ		& \qw	 	& \qw  \\ \\
}

&&\enspace

\Qcircuit @C=0.5em @R=0.2em @!R {
|q_1\rangle &&& \qw	&\ctrl{2}	&\qw		& \qw \\
|q_2\rangle &&& \gate{X}	&\ctrl{1}	&\gate{X}	& \qw   \\
|q_3\rangle &&& \gate{X}	&\gate{Z}	&\gate{X}	& \qw \\ \\
} \\
\hline

\bf{001} 
&&\enspace

\Qcircuit @C=0.5em @R=0.2em @!R {
|q_1\rangle &&& \gate{X}		&\ctrl{3}	& \gate{X}	& \qw 	 \\
|q_2\rangle &&& \gate{X}		&\ctrl{2}	& \gate{X}	& \qw 	 \\
|q_3\rangle &&& \qw		&\ctrl{1}	& \qw		& \qw 	 \\
|q_a\rangle &&& \qw 		&\targ		& \qw	 	& \qw  \\ \\
}

&&\enspace

\Qcircuit @C=0.5em @R=0.2em @!R {
|q_1\rangle &&& \gate{X}	&\ctrl{2}	&\gate{X}	& \qw \\
|q_2\rangle &&& \gate{X}	&\ctrl{1}	&\gate{X}	& \qw   \\
|q_3\rangle &&& \qw	&\gate{Z}	&\qw		& \qw \\ \\
}

&
\bf{101} 
&&\enspace

\Qcircuit @C=0.5em @R=0.2em @!R {
|q_1\rangle &&& \qw		&\ctrl{3}	& \qw		& \qw 	 \\
|q_2\rangle &&& \gate{X}		&\ctrl{2}	& \gate{X}	& \qw 	 \\
|q_3\rangle &&& \qw		&\ctrl{1}	& \qw	& \qw 	 \\
|q_a\rangle &&& \qw 		&\targ		& \qw 	& \qw  \\ \\
}

&&\enspace

\Qcircuit @C=0.5em @R=0.2em @!R {
|q_1\rangle &&& \qw	&\ctrl{2}	&\qw		& \qw \\
|q_2\rangle &&& \gate{X}	&\ctrl{1}	&\gate{X}	& \qw   \\
|q_3\rangle &&& \qw	&\gate{Z}	&\qw		& \qw \\ \\
} \\
\hline

\bf{010} 
&&\enspace

\Qcircuit @C=0.5em @R=0.2em @!R {
|q_1\rangle &&& \gate{X}		&\ctrl{3}	& \gate{X}	& \qw 	 \\
|q_2\rangle &&& \qw		&\ctrl{2}	& \qw		& \qw 	 \\
|q_3\rangle &&& \gate{X}		&\ctrl{1}	& \gate{X}	& \qw 	 \\
|q_a\rangle &&& \qw 		&\targ		& \qw	 	& \qw  \\ \\
}

&&\enspace

\Qcircuit @C=0.5em @R=0.2em @!R {
|q_1\rangle &&& \gate{X}	&\ctrl{2}	&\gate{X}	& \qw \\
|q_2\rangle &&& \qw	&\ctrl{1}	&\qw		& \qw   \\
|q_3\rangle &&& \gate{X}	&\gate{Z}	&\gate{X}	& \qw \\ \\
}

&
\bf{110} 
&&\enspace

\Qcircuit @C=0.5em @R=0.2em @!R {
|q_1\rangle &&& \qw		&\ctrl{3}	& \qw		& \qw 	 \\
|q_2\rangle &&& \qw		&\ctrl{2}	& \qw		& \qw 	 \\
|q_3\rangle &&& \gate{X}		&\ctrl{1}	& \gate{X}	& \qw 	 \\
|q_a\rangle &&& \qw 		&\targ		& \qw 		& \qw  \\ \\
}

&&\enspace

\Qcircuit @C=0.5em @R=0.2em @!R {
|q_1\rangle &&& \qw	&\ctrl{2}	&\qw		& \qw \\
|q_2\rangle &&& \qw	&\ctrl{1}	&\qw		& \qw   \\
|q_3\rangle &&& \gate{X}	&\gate{Z}	&\gate{X}	& \qw \\ \\
} \\
\hline

\bf{011} 
&&\enspace

\Qcircuit @C=0.5em @R=0.2em @!R {
|q_1\rangle &&& \gate{X}		&\ctrl{3}	& \gate{X}	& \qw 	 \\
|q_2\rangle &&& \qw		&\ctrl{2}	& \qw		& \qw 	 \\
|q_3\rangle &&& \qw		&\ctrl{1}	& \qw		& \qw 	 \\
|q_a\rangle &&& \qw 		&\targ		& \qw	 	& \qw  \\ \\
}

&&\enspace

\Qcircuit @C=0.5em @R=0.2em @!R {
|q_1\rangle &&& \gate{X}	&\ctrl{2}	&\gate{X}	& \qw \\
|q_2\rangle &&& \qw	&\ctrl{1}	&\qw		& \qw   \\
|q_3\rangle &&& \qw	&\gate{Z}	&\qw		& \qw \\ \\
}

&
\bf{111} 
&&\enspace

\Qcircuit @C=0.5em @R=0.65em @!R {
|q_1\rangle &&& \qw		&\ctrl{3}	& \qw		& \qw 	 \\
|q_2\rangle &&& \qw		&\ctrl{2}	& \qw		& \qw 	 \\
|q_3\rangle &&& \qw		&\ctrl{1}	& \qw		& \qw 	 \\
|q_a\rangle &&& \qw 		&\targ		& \qw 		& \qw  \\ \\
}

&&\enspace

\Qcircuit @C=0.5em @R=0.2em @!R {
|q_1\rangle &&& \qw	&\ctrl{2}	&\qw		& \qw \\
|q_2\rangle &&& \qw	&\ctrl{1}	&\qw		& \qw   \\
|q_3\rangle &&& \qw	&\gate{Z}	&\qw		& \qw \\ \\
} \\
\hline

\caption{Table of all oracles used for the single-solution Grover search algorithm.}
\label{tab:Grover1Oracles}
\end{longtable*}
\end{center}
}

\newcommand{\ChooseTwoOracles}{

\begin{center}
\begin{longtable*}{| p{1.1cm} | c c | c c || p{1.1cm} | c c | c c |}
\hline
Marked && Bit-Flip Oracle && Phase-Flip Oracle & Marked && Bit-Flip Oracle && Phase-Flip Oracle\\
\hline\hline
\endfirsthead
\multicolumn{4}{c}%
{\tablename\ \thetable\ -- \textit{Continued from previous page}} \\
\hline
Marked && Bit-Flip Oracle && Phase-Flip Oracle & Marked && Bit-Flip Oracle && Phase-Flip Oracle\\
\hline\hline
\endhead

\hline \multicolumn{4}{c}{\tablename\ \thetable\ -- \textit{Continued on next page}} \\ 
\endfoot

\endlastfoot

\bf{000, 001} 
&&\enspace
\Qcircuit @C=0.5em @R=0.2em @!R {
|q_1\rangle &&& \gate{X} 	& \ctrl{1} 		& \gate{X}	 	& \qw 	 \\
|q_2\rangle &&& \gate{X} 	& \ctrl{2} 		& \gate{X}		 & \qw 	 \\
|q_3\rangle &&& \qw		 & \qw 		& \qw		& \qw 	 \\
|q_a\rangle &&& \qw 		 & \targ		& \qw  		& \qw  \\ \\
} 
&&\enspace

\Qcircuit @C=0.5em @R=0.2em @!R {
|q_1\rangle &&& \gate{Z} & \ctrl{1} 	& \qw \\
|q_2\rangle &&& \gate{Z} & \gate{Z}	& \qw   \\
|q_3\rangle &&& \qw	  & \qw 		 & \qw \\ \\
} 

&

 \bf{010, 100} 
&&\enspace
\Qcircuit @C=0.5em @R=0.2em @!R {
 |q_1\rangle &&& \qw 		&\ctrl{1}	& \qw 		&\ctrl{1}	& \qw	 		& \qw 	 \\
 |q_2\rangle &&& \qw 		&\targ		& \ctrl{2} 	&\targ		& \qw			& \qw 	 \\
 |q_3\rangle &&& \gate{X}	&\qw		& \ctrl{1} 	&\qw		& \gate{X}		& \qw 	 \\
 |q_a\rangle &&& \qw 		&\qw		& \targ	&\qw		& \qw  		& \qw  \\ \\
} 
&&\enspace

\Qcircuit @C=0.5em @R=0.2em @!R {
|q_1\rangle &&& \gate{Z} 	& \ctrl{2} 	&\qw		& \qw \\
|q_2\rangle &&& \gate{Z} 	& \qw		&\ctrl{1}	& \qw   \\
|q_3\rangle &&& \qw	  	& \gate{Z}	&\gate{Z}	 & \qw \\ \\
}  \\

\hline

\bf{000, 010} 
&&\enspace

\Qcircuit @C=0.5em @R=0.2em @!R {
|q_1\rangle &&& \gate{X} 	& \ctrl{2} 		& \gate{X}	 	& \qw 	 \\
|q_2\rangle &&& \qw 		& \qw 		& \qw		 & \qw 	 \\
|q_3\rangle &&& \gate{X} 	& \ctrl{1} 		& \gate{X}		& \qw 	 \\
|q_a\rangle &&& \qw 		 & \targ		& \qw  		& \qw  \\ \\
}

&&\enspace

\Qcircuit @C=0.5em @R=0.2em @!R {
|q_1\rangle &&& \gate{Z} & \ctrl{2} 	& \qw \\
|q_2\rangle &&& \qw & \qw	& \qw   \\
|q_3\rangle &&& \gate{Z}	  & \gate{Z} 		 & \qw \\ \\
}

&
\bf{010, 101} 
&&\enspace

\Qcircuit @C=0.5em @R=0.2em @!R {
|q_1\rangle &&& \qw	& \ctrl{1}	& \ctrl{2}	&\qw	  	& \ctrl{2}	&\ctrl{1}	& \qw		& \qw 	 \\
|q_2\rangle &&& \qw	&\targ		& \qw		& \ctrl{2} 	& \qw		&\targ		& \qw		 & \qw 	 \\
|q_3\rangle &&& \gate{X}	&\qw		& \targ	& \ctrl{1} 	& \targ	&\qw		& \gate{X}	& \qw 	 \\
|q_a\rangle &&& \qw 	& \qw		& \qw	 	& \targ	& \qw		&\qw		& \qw  	& \qw  \\ \\
}

&&\enspace

\Qcircuit @C=0.5em @R=0.2em @!R {
|q_1\rangle &&& \qw	&\ctrl{1}	& \ctrl{2}	& \qw		& \qw \\
|q_2\rangle &&& \gate{Z}	&\gate{Z}	& \qw		&\ctrl{1}	& \qw   \\
|q_3\rangle &&& \qw	&\qw		& \gate{Z}	&\gate{Z}	& \qw \\ \\
} \\
\hline
\bf{000, 011} 
&&\enspace

\Qcircuit @C=0.5em @R=0.2em @!R {
|q_1\rangle &&& \gate{X}	& \qw	& \ctrl{2} 		& \qw	& \gate{X}	 	& \qw 	 \\
|q_2\rangle &&& \gate{X}	& \ctrl{1}	& \qw 		& \ctrl{1}	& \gate{X}		 & \qw 	 \\
|q_3\rangle &&& \qw		& \targ	& \ctrl{1} 		& \targ	& \qw		& \qw 	 \\
|q_a\rangle &&& \qw 		& \qw	 & \targ		& \qw	& \qw  		& \qw  \\ \\
}

&&\enspace

\Qcircuit @C=0.5em @R=0.2em @!R {
|q_1\rangle &&& \gate{Z}	&\ctrl{1}	& \ctrl{2}	& \qw \\
|q_2\rangle &&& \gate{Z}	&\gate{Z}	& \qw	& \qw   \\
|q_3\rangle &&& \gate{Z}	&\qw		& \gate{Z}	& \qw \\ \\
}

&
\bf{010, 110} 
&&\enspace

\Qcircuit @C=0.5em @R=0.2em @!R {
|q_1\rangle &&& \qw 		& \qw 			& \qw	 		& \qw 	 \\
|q_2\rangle &&& \qw 		& \ctrl{2} 		& \qw			 & \qw 	 \\
|q_3\rangle &&& \gate{X} 	& \ctrl{1} 		& \gate{X}		& \qw 	 \\
|q_a\rangle &&& \qw 		 & \targ		& \qw  		& \qw  \\ \\
}

&&\enspace

\Qcircuit @C=0.5em @R=0.2em @!R {
|q_1\rangle &&& \qw	 & \qw 		& \qw \\
|q_2\rangle &&& \gate{Z}	 & \ctrl{1}		& \qw   \\
|q_3\rangle &&& \qw	  & \gate{Z} 		 & \qw \\ \\
} \\
\hline

\bf{000, 100} 
&&\enspace
\Qcircuit @C=0.5em @R=0.2em @!R {
|q_1\rangle &&& \qw 	& \qw 		& \qw	 	& \qw 	 \\
|q_2\rangle &&& \gate{X} 	& \ctrl{2} 		& \gate{X}		 & \qw 	 \\
|q_3\rangle &&& \gate{X}	 & \ctrl{1} 		& \gate{X}		& \qw 	 \\
|q_a\rangle &&& \qw 	 & \targ		& \qw  		& \qw  \\ \\
} 
&&\enspace

\Qcircuit @C=0.5em @R=0.2em @!R {
|q_1\rangle &&& \qw 	& \qw 		& \qw \\
|q_2\rangle &&& \gate{Z} 	& \ctrl{1}		& \qw   \\
|q_3\rangle &&& \gate{Z}	 & \gate{Z} 	 & \qw \\ \\
} 

&

 \bf{010, 111} 
&&\enspace
\Qcircuit @C=0.5em @R=0.2em @!R {
 |q_1\rangle &&& \gate{X}	 	&\ctrl{2}	& \qw 	&\ctrl{2}	& \gate{X}		& \qw 	 \\
 |q_2\rangle &&& \qw 		&\qw		& \ctrl{2} 	&\qw		& \qw		& \qw 	 \\
 |q_3\rangle &&& \qw 		&\targ	& \ctrl{1} 	&\targ	& \qw		& \qw 	 \\
 |q_a\rangle &&& \qw 		&\qw		& \targ	&\qw		& \qw  		& \qw  \\ \\
} 
&&\enspace

\Qcircuit @C=0.5em @R=0.2em @!R {
|q_1\rangle &&& \qw 	& \ctrl{1} 	&\qw		& \qw \\
|q_2\rangle &&& \gate{Z} 	& \gate{Z}	&\ctrl{1}	& \qw   \\
|q_3\rangle &&& \qw	  	& \qw	&\gate{Z}	 & \qw \\ \\
}  \\

\hline

\bf{000, 101} 
&&\enspace
\Qcircuit @C=0.5em @R=0.2em @!R {
 |q_1\rangle &&& \gate{X} 	&\ctrl{2}	& \qw 		&\ctrl{2}	& \gate{X}	 	& \qw 	 \\
 |q_2\rangle &&& \gate{X} 	&\qw		& \ctrl{2} 	&\qw		& \gate{X}		& \qw 	 \\
 |q_3\rangle &&& \qw		&\targ		& \ctrl{1} 	&\targ		& \qw			& \qw 	 \\
 |q_a\rangle &&& \qw 		&\qw		& \targ	&\qw		& \qw  		& \qw  \\ \\
} 
&&\enspace

\Qcircuit @C=0.5em @R=0.2em @!R {
|q_1\rangle &&& \gate{Z} 	& \ctrl{1} 	&\qw		& \qw \\
|q_2\rangle &&& \gate{Z} 	& \gate{Z}	&\ctrl{1}	& \qw   \\
|q_3\rangle &&& \gate{Z}	  	& \qw		&\gate{Z}	 & \qw \\ \\
} 

&

 \bf{011, 100} 
&&\enspace
\Qcircuit @C=0.5em @R=0.65em @!R {
|q_1\rangle &&&  \ctrl{1}		& \ctrl{2}	&\qw	  	& \ctrl{2}	&\ctrl{1}	& \qw 	 \\
|q_2\rangle &&& \targ		& \qw		& \ctrl{2} 	& \qw		&\targ		 & \qw 	 \\
|q_3\rangle &&& \qw		& \targ	& \ctrl{1} 	& \targ	&\qw		& \qw 	 \\
|q_a\rangle &&& \qw		& \qw	 	& \targ	& \qw		&\qw		& \qw  \\ \\
}

&&\enspace

\Qcircuit @C=0.5em @R=0.2em @!R {
|q_1\rangle &&& \gate{Z}	&\ctrl{1}	& \ctrl{2}	& \qw		& \qw \\
|q_2\rangle &&& \qw	&\gate{Z}	& \qw		&\ctrl{1}	& \qw   \\
|q_3\rangle &&& \qw	&\qw		& \gate{Z}	&\gate{Z}	& \qw \\ \\
}  \\
\hline

\bf{000, 110} 
&&\enspace
\Qcircuit @C=0.5em @R=0.2em @!R {
 |q_1\rangle &&& \gate{X} 	&\ctrl{1}	& \qw 		&\ctrl{1}	& \gate{X}	 		& \qw 	 \\
 |q_2\rangle &&& \qw 		&\targ		& \ctrl{2} 	&\targ		& \qw			& \qw 	 \\
 |q_3\rangle &&& \gate{X}	&\qw		& \ctrl{1} 	&\qw		& \gate{X}		& \qw 	 \\
 |q_a\rangle &&& \qw 		&\qw		& \targ	&\qw		& \qw  		& \qw  \\ \\
} 
&&\enspace

\Qcircuit @C=0.5em @R=0.2em @!R {
|q_1\rangle &&& \gate{Z} 	& \ctrl{2} 	&\qw		& \qw \\
|q_2\rangle &&& \gate{Z} 	& \qw		&\ctrl{1}	& \qw   \\
|q_3\rangle &&& \gate{Z}	  	& \gate{Z}	&\gate{Z}	 & \qw \\ \\
}

&

 \bf{011, 101} 
&&\enspace
\Qcircuit @C=0.5em @R=0.65em @!R {
 |q_1\rangle &&& \ctrl{1}		& \qw 		&\ctrl{1}	& \qw 	 \\
 |q_2\rangle &&& \targ		& \ctrl{2} 	&\targ		& \qw 	 \\
 |q_3\rangle &&& \qw		& \ctrl{1} 	&\qw		& \qw 	 \\
 |q_a\rangle &&& \qw		& \targ	&\qw		& \qw  \\ \\
} 
&&\enspace

\Qcircuit @C=0.5em @R=0.2em @!R {
|q_1\rangle &&& \ctrl{2} 	&\qw		& \qw \\
|q_2\rangle &&& \qw	&\ctrl{1}	& \qw   \\
|q_3\rangle &&& \gate{Z}	&\gate{Z}	 & \qw \\ \\
}  \\

\hline

\bf{000, 111} 
&&\enspace

\Qcircuit @C=0.5em @R=0.2em @!R {
|q_1\rangle &&& \gate{X}	& \ctrl{1}	& \ctrl{2}	&\qw	  	& \ctrl{2}	&\ctrl{1}	& \gate{X}	& \qw 	 \\
|q_2\rangle &&& \qw	&\targ		& \qw		& \ctrl{2} 	& \qw		&\targ		& \qw		 & \qw 	 \\
|q_3\rangle &&& \qw	&\qw		& \targ	& \ctrl{1} 	& \targ	&\qw		& \qw		& \qw 	 \\
|q_a\rangle &&& \qw 	& \qw		& \qw	 	& \targ	& \qw		&\qw		& \qw  	& \qw  \\ \\
}

&&\enspace

\Qcircuit @C=0.5em @R=0.2em @!R {
|q_1\rangle &&& \gate{Z}	&\ctrl{1}	& \ctrl{2}	& \qw		& \qw \\
|q_2\rangle &&& \gate{Z}	&\gate{Z}	& \qw		&\ctrl{1}	& \qw   \\
|q_3\rangle &&& \gate{Z}	&\qw		& \gate{Z}	&\gate{Z}	& \qw \\ \\
}

&
\bf{011, 110} 
&&\enspace

\Qcircuit @C=0.5em @R=0.65em  @!R{
|q_1\rangle &&& \ctrl{2}	&\qw	  	& \ctrl{2}	& \qw 	 \\
|q_2\rangle &&& \qw	& \ctrl{2} 	& \qw		& \qw 	 \\
|q_3\rangle &&& \targ	& \ctrl{1} 	& \targ	& \qw 	 \\
|q_a\rangle &&& \qw	& \targ	& \qw		& \qw  \\ \\
}

&&\enspace

\Qcircuit @C=0.5em @R=0.2em @!R {
|q_1\rangle &&&\ctrl{1}	& \qw		& \qw \\
|q_2\rangle &&&\gate{Z}	&\ctrl{1}	& \qw   \\
|q_3\rangle &&&\qw	&\gate{Z}	& \qw \\ \\
} \\
\hline

\bf{001, 010} 
&&\enspace

\Qcircuit @C=0.5em @R=0.2em @!R {
|q_1\rangle &&& \gate{X}	& \qw		& \ctrl{3} 	& \qw		& \gate{X}	 	& \qw 	 \\
|q_2\rangle &&& \qw	& \ctrl{1}	& \qw 		& \ctrl{1}	& \qw			 & \qw 	 \\
|q_3\rangle &&& \qw	& \targ	& \ctrl{1} 	& \targ	& \qw			& \qw 	 \\
|q_a\rangle &&& \qw 	& \qw	 	& \targ	& \qw		& \qw  		& \qw  \\ \\
}

&&\enspace

\Qcircuit @C=0.5em @R=0.2em @!R {
|q_1\rangle &&& \qw	&\ctrl{1}	& \ctrl{2}	& \qw \\
|q_2\rangle &&& \gate{Z}	&\gate{Z}	& \qw		& \qw   \\
|q_3\rangle &&& \gate{Z}	&\qw		& \gate{Z}	& \qw \\ \\
}

&
\bf{011, 111} 
&&\enspace

\Qcircuit @C=0.5em @R=0.65em @!R {
|q_1\rangle &&& 		 \qw 				 	& \qw 	 \\
|q_2\rangle &&&  		 \ctrl{2} 				 & \qw 	 \\
|q_3\rangle &&&  		 \ctrl{1} 				& \qw 	 \\
|q_a\rangle &&&  		  \targ		 		& \qw  \\ \\
}

&&\enspace

\Qcircuit @C=0.5em @R=0.2em @!R {
|q_1\rangle &&&  \qw 		& \qw \\
|q_2\rangle &&&  \ctrl{1}		& \qw   \\
|q_3\rangle &&&  \gate{Z} 		 & \qw \\ \\
} \\
\hline

\bf{001, 011} 
&&\enspace

\Qcircuit @C=0.5em @R=0.2em @!R {
|q_1\rangle &&& \gate{X} 	& \ctrl{2} 		& \gate{X}	 	& \qw 	 \\
|q_2\rangle &&& \qw 		& \qw 			& \qw			 & \qw 	 \\
|q_3\rangle &&& \qw	 	& \ctrl{1} 		& \qw			& \qw 	 \\
|q_a\rangle &&& \qw 		 & \targ		& \qw  		& \qw  \\ \\
}

&&\enspace

\Qcircuit @C=0.5em @R=0.2em @!R {
|q_1\rangle &&& \qw 		& \ctrl{2} 	& \qw \\
|q_2\rangle &&& \qw 		& \qw		& \qw   \\
|q_3\rangle &&& \gate{Z}	  	& \gate{Z} 	& \qw \\ \\
}

&
\bf{100, 101} 
&&\enspace

\Qcircuit @C=0.5em @R=0.2em @!R {
|q_1\rangle &&& \qw	&\ctrl{3}	& \qw		& \qw 	 \\
|q_2\rangle &&& \gate{X} & \ctrl{2} 	& \gate{X}	 & \qw 	 \\
|q_3\rangle &&& \qw	& \qw 		& \qw		& \qw 	 \\
|q_a\rangle &&& \qw 	& \targ	& \qw  	& \qw  \\ \\
}

&&\enspace

\Qcircuit @C=0.5em @R=0.2em @!R {
|q_1\rangle &&& \gate{Z}	&\ctrl{1}	& \qw \\
|q_2\rangle &&& \qw	&\gate{Z}	& \qw   \\
|q_3\rangle &&& \qw	&\qw		& \qw \\ \\
} \\
\hline

\bf{001, 100} 
&&\enspace
\Qcircuit @C=0.5em @R=0.2em @!R {
 |q_1\rangle &&& \qw 		&\ctrl{2}	& \qw 		&\ctrl{2}	& \qw		 	& \qw 	 \\
 |q_2\rangle &&& \gate{X} 	&\qw		& \ctrl{2} 	&\qw		& \gate{X}		& \qw 	 \\
 |q_3\rangle &&& \qw		&\targ		& \ctrl{1} 	&\targ		& \qw			& \qw 	 \\
 |q_a\rangle &&& \qw 		&\qw		& \targ	&\qw		& \qw  		& \qw  \\ \\
} 
&&\enspace

\Qcircuit @C=0.5em @R=0.2em @!R {
|q_1\rangle &&& \gate{Z} 	& \ctrl{1} 	&\qw		& \qw \\
|q_2\rangle &&& \qw	 	& \gate{Z}	&\ctrl{1}	& \qw   \\
|q_3\rangle &&& \gate{Z}	  	& \qw		&\gate{Z}	 & \qw \\ \\
} 

&

 \bf{100, 110} 
&&\enspace
\Qcircuit @C=0.5em @R=0.2em @!R {
|q_1\rangle &&&  \qw		&\ctrl{3}	&\qw		& \qw 	 \\
|q_2\rangle &&& \qw		& \qw 		&\qw		 & \qw 	 \\
|q_3\rangle &&& \gate{X}		& \ctrl{1} 	&\gate{X}	& \qw 	 \\
|q_a\rangle &&& \qw		& \targ	&\qw		& \qw  \\ \\
}

&&\enspace

\Qcircuit @C=0.5em @R=0.2em @!R {
|q_1\rangle &&& \gate{Z}	& \ctrl{2}	& \qw \\
|q_2\rangle &&& \qw	& \qw		& \qw   \\
|q_3\rangle &&& \qw	& \gate{Z}	& \qw \\ \\
}  \\
\hline

\bf{001, 101} 
&&\enspace
\Qcircuit @C=0.5em @R=0.2em @!R {
|q_1\rangle &&& \qw 	& \qw 			& \qw	 	& \qw 	 \\
|q_2\rangle &&& \gate{X} & \ctrl{2} 		& \gate{X}	 & \qw 	 \\
|q_3\rangle &&& \qw	 & \ctrl{1} 		& \qw		& \qw 	 \\
|q_a\rangle &&& \qw 	 & \targ		& \qw  	& \qw  \\ \\
} 
&&\enspace

\Qcircuit @C=0.5em @R=0.2em @!R {
|q_1\rangle &&& \qw 	& \qw 		& \qw \\
|q_2\rangle &&& \qw	& \ctrl{1}	& \qw   \\
|q_3\rangle &&& \gate{Z}	 & \gate{Z} 	 & \qw \\ \\
} 

&

 \bf{100, 111} 
&&\enspace
\Qcircuit @C=0.5em @R=0.2em @!R {
 |q_1\rangle &&& \qw	 	&\qw		& \ctrl{3} 	&\qw		& \qw		& \qw 	 \\
 |q_2\rangle &&& \gate{X} 	&\ctrl{1}	& \qw 		&\ctrl{1}	& \gate{X}	& \qw 	 \\
 |q_3\rangle &&& \qw 		&\targ		& \ctrl{1} 	&\targ		& \qw		& \qw 	 \\
 |q_a\rangle &&& \qw 		&\qw		& \targ	&\qw		& \qw  	& \qw  \\ \\
} 
&&\enspace

\Qcircuit @C=0.5em @R=0.2em @!R {
|q_1\rangle &&& \gate{Z} 	& \ctrl{1} 	&\ctrl{2}	& \qw \\
|q_2\rangle &&& \qw 		& \gate{Z}	&\qw		& \qw   \\
|q_3\rangle &&& \qw	  	& \qw		&\gate{Z}	 & \qw \\ \\
}  \\

\hline

\bf{001, 110} 
&&\enspace

\Qcircuit @C=0.5em @R=0.2em @!R {
|q_1\rangle &&& \qw		& \ctrl{1}	& \ctrl{2}	&\qw	  	& \ctrl{2}	&\ctrl{1}	& \qw		& \qw 	 \\
|q_2\rangle &&& \gate{X}	&\targ	& \qw	& \ctrl{2} 	& \qw	&\targ	& \gate{X}		 & \qw 	 \\
|q_3\rangle &&& \qw		&\qw		& \targ	& \ctrl{1} 	& \targ	&\qw		& \qw		& \qw 	 \\
|q_a\rangle &&& \qw 	& \qw	& \qw	 & \targ	& \qw	&\qw		& \qw  		& \qw  \\ \\
}

&&\enspace

\Qcircuit @C=0.5em @R=0.2em @!R {
|q_1\rangle &&& \qw		&\ctrl{1}	& \ctrl{2}	& \qw	& \qw \\
|q_2\rangle &&& \qw		&\gate{Z}	& \qw	&\ctrl{1}	& \qw   \\
|q_3\rangle &&& \gate{Z}	&\qw		& \gate{Z}	 &\gate{Z}	& \qw \\ \\
}

&
\bf{101, 110} 
&&\enspace

\Qcircuit @C=0.5em @R=0.65em  @!R{
|q_1\rangle &&& \qw		&\ctrl{3}	 & \qw	& \qw 	 \\
|q_2\rangle &&& \ctrl{1}	& \qw 	& \ctrl{1}	& \qw 	 \\
|q_3\rangle &&& \targ	& \ctrl{1} 	& \targ	& \qw 	 \\
|q_a\rangle &&& \qw		& \targ	& \qw	& \qw  \\ \\
}

&&\enspace

\Qcircuit @C=0.5em @R=0.2em @!R {
|q_1\rangle &&&\ctrl{1}	& \ctrl{2}	& \qw \\
|q_2\rangle &&&\gate{Z}	&\qw		& \qw   \\
|q_3\rangle &&&\qw		&\gate{Z}	& \qw \\ \\
} \\
\hline

\bf{001, 111} 
&&\enspace
\Qcircuit @C=0.5em @R=0.2em @!R {
 |q_1\rangle &&& \gate{X} 	&\ctrl{1}	& \qw 		&\ctrl{1}	& \gate{X}	 	& \qw 	 \\
 |q_2\rangle &&& \qw 		&\targ		& \ctrl{2} 	&\targ		& \qw			& \qw 	 \\
 |q_3\rangle &&& \qw		&\qw		& \ctrl{1} 	&\qw		& \qw			& \qw 	 \\
 |q_a\rangle &&& \qw 		&\qw		& \targ	&\qw		& \qw  		& \qw  \\ \\
} 
&&\enspace

\Qcircuit @C=0.5em @R=0.2em @!R {
|q_1\rangle &&& \qw	 	& \ctrl{2} 	&\qw		& \qw \\
|q_2\rangle &&& \qw	 	& \qw		&\ctrl{1}	& \qw   \\
|q_3\rangle &&& \gate{Z}	  	& \gate{Z}	&\gate{Z}	 & \qw \\ \\
}

&

 \bf{101, 111} 
&&\enspace
\Qcircuit @C=0.5em @R=0.65em @!R {
 |q_1\rangle &&&  \ctrl{3} 		& \qw 	 \\
 |q_2\rangle &&&  \qw 		& \qw 	 \\
 |q_3\rangle &&&  \ctrl{1} 		& \qw 	 \\
 |q_a\rangle &&&  \targ		& \qw  \\ \\
} 
&&\enspace

\Qcircuit @C=0.5em @R=0.2em @!R {
|q_1\rangle &&& \ctrl{2} 	& \qw \\
|q_2\rangle &&& \qw	& \qw   \\
|q_3\rangle &&& \gate{Z}	 & \qw \\ \\
}  \\

\hline

\bf{010, 011} 
&&\enspace
\Qcircuit @C=0.5em @R=0.2em @!R {
 |q_1\rangle &&& \gate{X} 	& \ctrl{3} 	& \gate{X}	 	& \qw 	 \\
 |q_2\rangle &&& \qw 		& \ctrl{2} 	& \qw			& \qw 	 \\
 |q_3\rangle &&& \qw		& \qw	 	& \qw			& \qw 	 \\
 |q_a\rangle &&& \qw 		& \targ	& \qw  		& \qw  \\ \\
} 
&&\enspace

\Qcircuit @C=0.5em @R=0.2em @!R {
|q_1\rangle &&& \qw	 	& \ctrl{1} 	& \qw \\
|q_2\rangle &&& \gate{Z}	 	& \gate{Z}	& \qw   \\
|q_3\rangle &&& \qw	  	& \qw		 & \qw \\ \\
}

&

 \bf{110, 111} 
&&\enspace
\Qcircuit @C=0.5em @R=0.65em @!R {
 |q_1\rangle &&&  \ctrl{3} 		& \qw 	 \\
 |q_2\rangle &&&  \ctrl{2} 		& \qw 	 \\
 |q_3\rangle &&&  \qw 		& \qw 	 \\
 |q_a\rangle &&&  \targ		& \qw  \\ \\
} 
&&\enspace

\Qcircuit @C=0.5em @R=0.2em @!R {
|q_1\rangle &&& \ctrl{1} 	& \qw \\
|q_2\rangle &&& \gate{Z}	& \qw   \\
|q_3\rangle &&& \qw	 & \qw \\ \\
}  \\

\hline

\caption{Table of all oracles used for the two-solution Grover search algorithm.}
\label{tab:Grover2Oracles}
\end{longtable*}
\end{center}
}

\begin{document}


\title{Complete 3-Qubit Grover Search on a Programmable Quantum Computer}



\author{C. Figgatt}
\affiliation{Joint Quantum Institute, Department of Physics, and Joint Center for Quantum Information and Computer Science, University of Maryland, College Park, MD 20742, USA}
\author{D. Maslov}
\affiliation{National Science Foundation, Arlington, VA 22230, USA}
\affiliation{Joint Quantum Institute, Department of Physics, and Joint Center for Quantum Information and Computer Science, University of Maryland, College Park, MD 20742, USA}
\author{K. A. Landsman}
\author{N. M. Linke}
\author{S. Debnath}
\affiliation{Joint Quantum Institute, Department of Physics, and Joint Center for Quantum Information and Computer Science, University of Maryland, College Park, MD 20742, USA}
\author{C. Monroe}
\affiliation{Joint Quantum Institute, Department of Physics, and Joint Center for Quantum Information and Computer Science, University of Maryland, College Park, MD 20742, USA}
\affiliation{IonQ Inc., College Park, MD 20742, USA}

\date{\today}

\pacs{}

\maketitle

\textbf{
Searching large databases is an important problem with broad applications. The Grover search algorithm \cite{GroverOriginal97, BoyerGrover98} provides a powerful method for quantum computers to perform searches with a quadratic speedup in the number of required database queries over classical computers. 
It is an optimal search algorithm for a quantum computer \cite{BennettBounds97}, and has further applications as a subroutine for other quantum algorithms \cite{MagniezGroverSubroutine07, DurrGroverSubroutine06}.
Searches with two qubits have been demonstrated on a variety of platforms \cite{ChuangNMRGrover98, BhattacharyaFourierOptics02, BrickmanIons05, WaltherPhotonClusters05, DicarloSC09, BarzPhotonics12} and proposed for others \cite{MolmerRydProposal11}, 
but larger search spaces have only been demonstrated on a non-scalable NMR system \cite{Vandersypen3qubitNMR00}.
Here, we report results for a complete three-qubit Grover search algorithm using the scalable quantum computing technology of trapped atomic ions \cite{Debnath16}, with better-than-classical performance. 
The algorithm is performed for all 8 possible single-result oracles and all 28 possible two-result oracles. Two methods of state marking are used for the oracles: a phase-flip method employed by other experimental demonstrations, and a Boolean method requiring an ancilla qubit that is directly equivalent to the state-marking scheme required to perform a classical search. All quantum solutions are shown to outperform their classical counterparts. 
We also report the first implementation of a Toffoli-4 gate, which is used along with Toffoli-3 gates to construct the algorithms; these gates have process fidelities of 70.5\% and 89.6\%, respectively. 
}

\ConceptFigure

The Grover search algorithm has 4 stages: initialization, oracle, amplification, and measurement, as shown in Figure \ref{fig:GroverConcept}(a). The initialization stage creates an equal superposition of all states. The oracle stage marks the solution(s) by flipping the sign of that state's amplitude. The amplification stage performs a reflection about the mean, thus increasing the amplitude of the marked state. Finally, the algorithm output is measured. For a search database of size $N$, the single-shot probability of measuring the correct answer is maximized to near-unity by repeating the oracle and amplification stages $O(\sqrt{N})$ times \cite{GroverOriginal97, BoyerGrover98}.  
By comparison, a classical search algorithm will get the correct answer after an average of $N/2$ queries of the oracle. For large databases, this quadratic speedup represents a significant advantage for quantum computers.

Here, we perform the Grover search algorithm on $n=3$ qubits, which corresponds to a search database of size $N=2^n=8$. All searches are performed with a single iteration. For a single-solution algorithm ($t=1$), the algorithmic probability of measuring the correct state after one iteration is $t\cdot\left( \left[ \frac{N-2t}{N}  + \frac{2(N-t)}{N} \right] \frac{1}{\sqrt{N}} \right)^2=\left( \frac{5}{4\sqrt{2}} \right)^2 = 78.125\%$ \cite{BoyerGrover98}, compared to $\frac{t}{N}+\frac{N-t}{N}\cdot\frac{t}{N-1} = \frac{1}{8}+ \frac{7}{8}\cdot\frac{1}{7} =  25\%$ for the optimal classical search strategy, which consists of a single query followed by a random guess in the event the query failed. 
In the two-solution case ($t=2$), where two states are marked as correct answers during the oracle stage and both states' amplitudes are amplified in the algorithm's amplification stage, the probability of measuring one of the two correct answers is $100\%$ for the quantum case, as compared to $\frac{13}{28} \approx 46.4\%$ for the classical case.

We examine two alternative methods of encoding the marked state within the oracle. While both methods are mathematically equivalent \cite{NielsenChuang11}, only one is directly comparable to a classical search. The previously-undemonstrated Boolean method requires the use of an ancilla qubit initialized to $|1\rangle$, as shown in Figure \ref{fig:GroverConcept}(b). The oracle is determined by constructing a circuit out of $NOT$ and $C^k(NOT)$ ($k\leq n$) gates such that, were the oracle circuit to be implemented classically, the ancilla bit would flip if and only if the input to the circuit is one of the marked states. By using classically available gates, this oracle formulation is directly equivalent to the classical search algorithm, and therefore can most convincingly demonstrate the quantum algorithm's superiority. On a quantum computer, because the initialization sets up an equal superposition of all possible input states, the $C^{n}(NOT)$ gate targeted on the ancilla provides a phase kickback that flips the phase of the marked state(s) in the data qubits. An example oracle is shown in Figure \ref{fig:GroverConcept}(c) to illustrate this. The phase method of oracle implementation does not require the ancilla qubit. Instead, the oracle is implemented with a circuit consisting of $Z$ and $C^k(Z)$ ($k \leq n-1$) gates that directly flip the phase(s) of the state(s) to be marked (see Figures \ref{fig:GroverConcept}(d-e)). 

The experiments presented here were performed on a programmable quantum computer consisting of a linear chain of five trapped $^{171}$Yb$^+$ ions \cite{Milburn00,Olmschenk07} that are laser cooled near the motional ground state. Qubits are comprised of the first-order magnetic-field-insensitive pair of clock states in the hyperfine-split $^2S_{1/2}$ manifold, with $|0\rangle \equiv |F=0; m_F=0\rangle$ and $|1\rangle \equiv |F=1; m_F=0\rangle$ having a 12.642821 GHz frequency difference. Optical pumping initializes all qubits to the $|0\rangle$ state. We execute modular one- and two-qubit gates through Raman transitions driven by a beat note between counter-propagating beams from a pulsed laser \cite{HayesCombs10}, which couples the qubit transition to the collective transverse modes of motion of the ion chain. The qubit-motion interaction provides entangling two-qubit Ising gates \cite{Milburn00,Solano99,Molmer99}. A pulse-segmentation scheme modulates the amplitude and phase of the Raman laser to drive high-fidelity entangling gates using all modes of motion \cite{ZhuGates06,Choi14}. Individual optical addressing of each ion with one Raman beam provides arbitrary single-qubit rotations ($R(\theta,\phi)$) as well as gates between arbitrary pairs of ions ($XX(\chi)$) (see Supplmentary Materials for details). State-dependent fluorescence detection with each ion mapped to a separate PMT channel allows for individual ion readout \cite{Debnath16}.

\begin{figure}
\centering
\begin{tabular}[c]{c }
\multicolumn{1}{l}{\bf{(a) Toffoli-3 Data}} \\
\multicolumn{1}{c}{\includegraphics[width=0.75\columnwidth,valign=T]{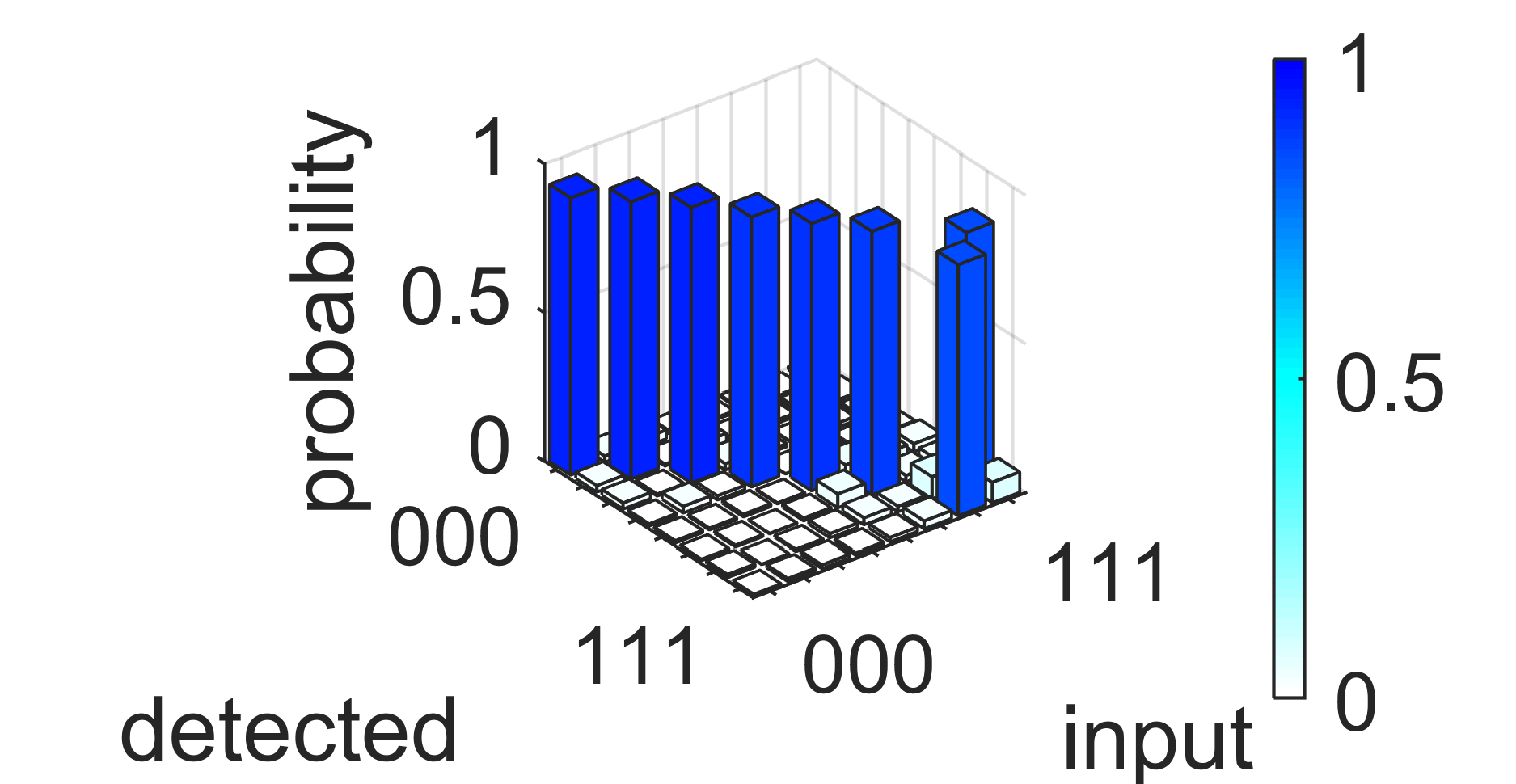}}\\  
\\
\multicolumn{1}{l}{\bf{(b) Toffoli-3 Abbreviated Circuit}} \\
\\
\multicolumn{1}{c}{\ToffCompressedCircuit} \\
\\
\multicolumn{1}{l}{\bf{(c) Toffoli-4 Data}} \\
\multicolumn{1}{c}{\includegraphics[width=0.99\columnwidth,valign=T]{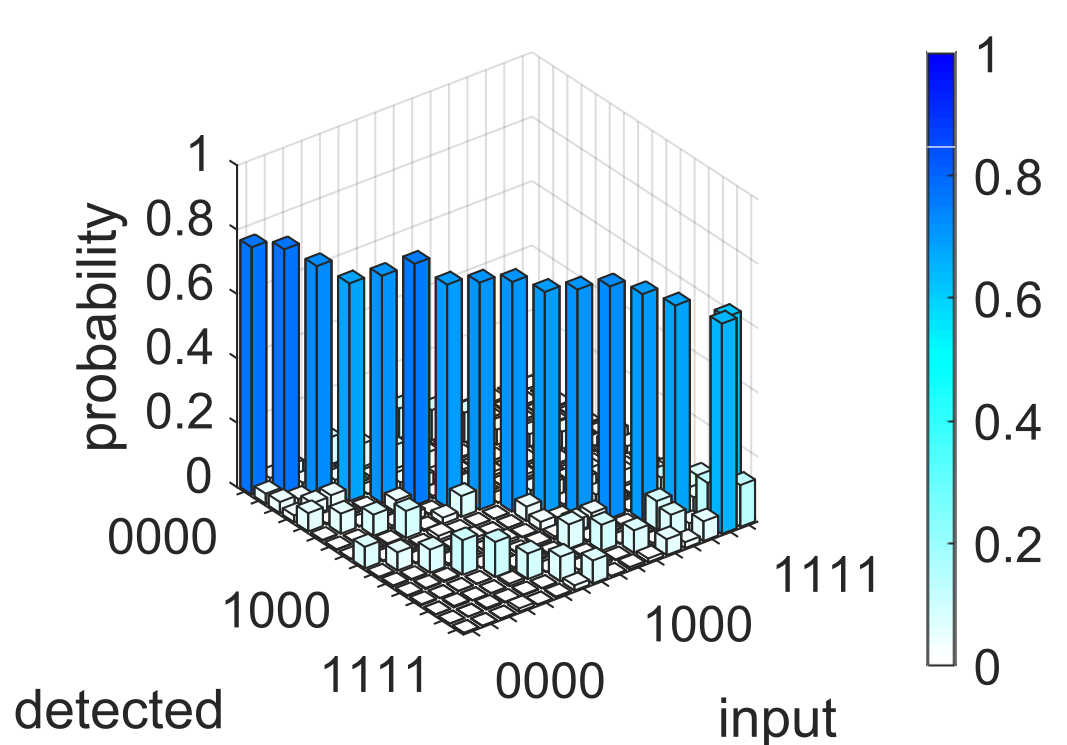}}
\end{tabular}
\caption{(a) Measured truth table for a Toffoli-3 gate. The average process fidelity is 89.6(2)\%, corrected for a 1.5\% average state preparation and measurement (SPAM) error. (b) Abbreviated circuit for implementing Toffoli-3 (see Supplementary Materials for details). (c) Measured truth table for a Toffoli-4 gate performed with 3 controls, 1 target, and 1 ancilla qubit. The average process fidelity is 70.5(3)\%, corrected for a 1.9\% average SPAM error.}
\label{fig:Toffoli}
\end{figure}

Successful demonstration of the Grover search algorithm first requires the implementation of its subroutines. Controlled-NOT ($CNOT$) gates constructed from an $XX\left(\frac{\pi}{4}\right)$ gate and single-qubit rotations have been demonstrated on this system previously \cite{Debnath16}. Here, we show results for a controlled-controlled-NOT ($C^2(NOT)$), or Toffoli-3, gate, with a process fidelity of 89.6(2)\% (see Figure \ref{fig:Toffoli}(a)). Toffoli-3 gates have been previously performed in NMR systems \cite{CoryNMRToffoli98} and ion traps \cite{Monz09}, including this system \cite{LinkeComparison17}. We employed a limited tomography procedure to verify that the Toffoli-3 gate performed had no spurious phases on the outputs (see Supplementary Materials).

Our Toffoli-3 is constructed from 5 two-qubit gates (three $XX\left(\frac{\pi}{8}\right)$ and two $XX\left(\frac{\pi}{4}\right)$ gates) in a manner similar to the Toffoli gate demonstrated in \cite{Vandersypen3qubitNMR00}. Any doubly-controlled unitary $C^2(U)$ operation can be performed with 5 two-qubit interactions (two $CNOT$s, two $C(V)$s, and one $C(V^{\dagger})$) if a controlled-$V$ operation is available such that $V^2=U$ \cite{Barenco95}. Since $\left[ XX\left(\frac{\pi}{8}\right) \right]^2 = XX\left(\frac{\pi}{4}\right)$, we can add single-qubit rotations to construct a Toffoli-3 gate with minimal use of two-qubit gates, as shown in Figure \ref{fig:Toffoli}(b) (see Supplementary Materials for a detailed circuit diagram). This compares favorably to the 6 two-qubit gates that would be necessary if only $CNOT$ (or equivalently, $XX\left(\frac{\pi}{4}\right)$) gates were available. These constructions also provide for the implementation of $C^2(Z)$ and $C(Z)$ gates, which can be constructed by adding a few single-qubit rotations to a Toffoli-3 or $CNOT$ gate, respectively (see Supplementary Material for circuits). For all circuits, the single-qubit rotations are further optimized to minimize total rotation time \cite{MaslovCircuits17}. 

We use a related strategy to construct a Toffoli-4 gate, and report an average process fidelity of 70.5(3)\% (see Figure \ref{fig:Toffoli}(c)). Using the methods described in \cite{maslovToffoli16}, we construct a circuit with 3 control qubits, 1 target, and 1 ancilla qubit, requiring 11 two-qubit gates (see Supplementary Materials for circuit). By again using both $XX\left(\frac{\pi}{4}\right)$ and $XX\left(\frac{\pi}{8}\right)$ gates, we are able to save one two-qubit gate relative to a construction limited to $CNOT$ gates \cite{maslovToffoli16}.

\begin{figure*}
\includegraphics[width=\textwidth]{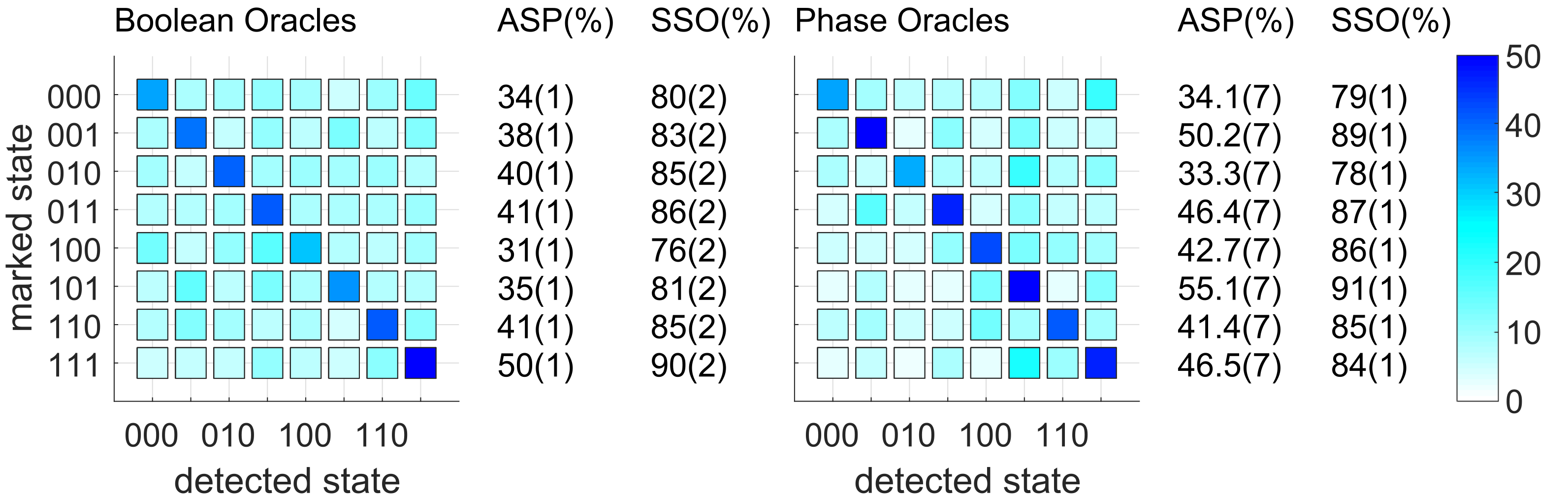}\\
\caption{Results from a single iteration of a single-solution Grover search algorithm performed on a 3-qubit database. Data for the Boolean oracle formulation is shown on the left, and data for the phase oracle formulation is shown on the right. The plots show the probability of detecting each output state. All values shown are percents, with a theoretical ASP of 78.1\% and theoretical SSO of 100\%. Data is corrected for average SPAM errors of 1\%.}
\label{fig:choose1}
\end{figure*}

\begin{figure*}
\includegraphics[width=\textwidth]{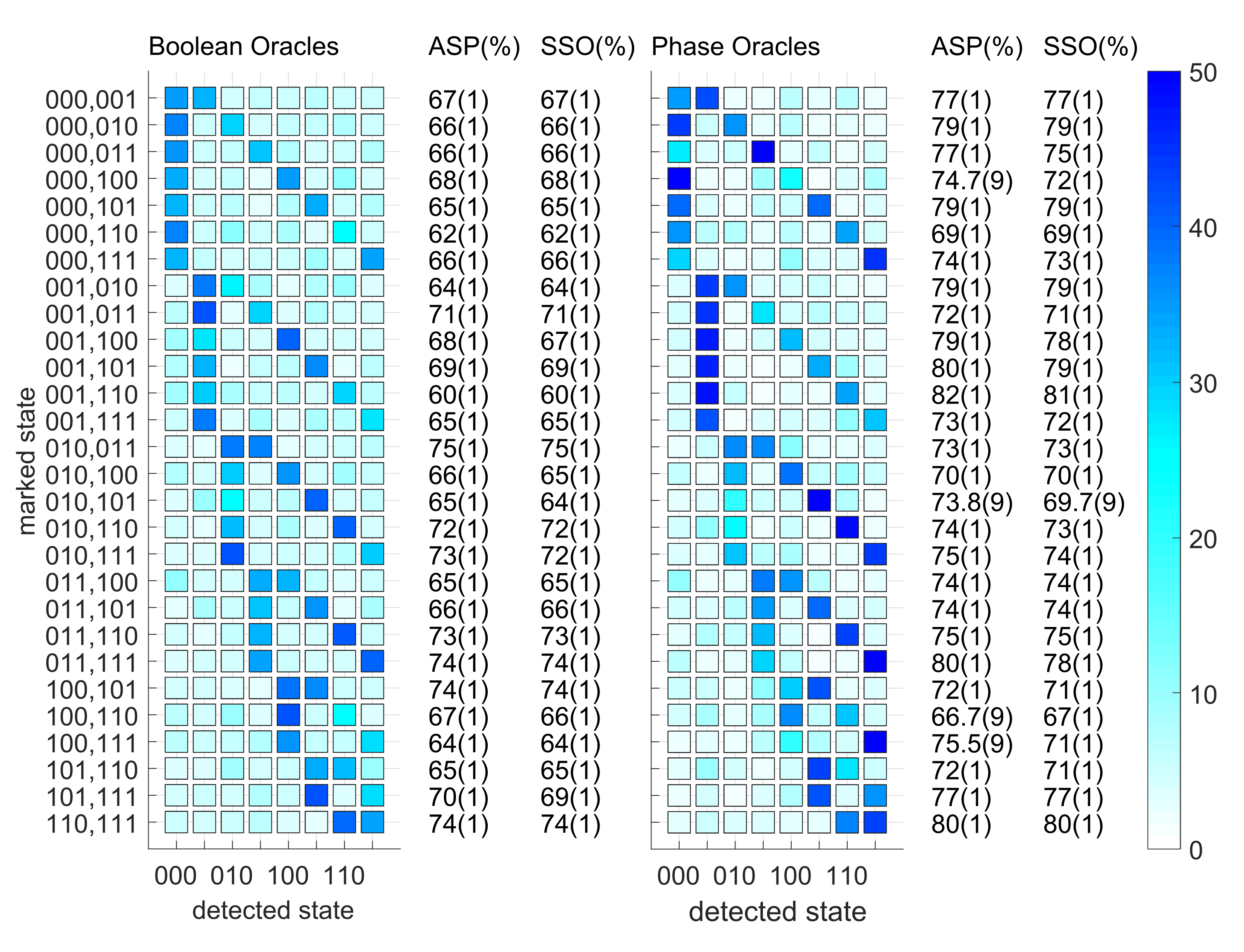}
\caption{Results from the execution of a two-solution Grover search algorithm performed on a 3-qubit database. Data for the Boolean oracle formulation is shown on the left, and data for the phase oracle formulation is shown on the right. The plots show the probability of detecting each output state. All values shown are percents. The ASP is the sum of the probabilities of detecting each of the two marked states. Data is corrected for average SPAM errors of 1\%.}
\label{fig:choose2}
\end{figure*}

Figures \ref{fig:choose1} and \ref{fig:choose2} show the results, respectively, of single- and two-solution Grover search algorithms, each using both the Boolean and phase marking methods. All possible oracles are tested to demonstrate a complete Grover search (see Supplementary Materials). Two figures of merit are provided with the data for each oracle. The algorithm success probability (ASP) is the probability of measuring the marked state as the experimental outcome. For the two-solution algorithm, the ASP is calculated by summing the probabilities of measuring each of the two marked states. The squared statistical overlap (SSO) measures the statistical overlap between the measured and expected populations for all states: SSO $= \left( \sum_{j=0}^N \sqrt{e_j m_j} \right)^2$, where $e_j$ is the expected population and $m_j$ is the measured population for each state $j$ \cite{ChiaveriniSSO05}. Additionally, all of the data shown in this paper is corrected to account for state preparation and measurement (SPAM) errors (see figure captions for values), similar to the method proposed in \cite{ShenSPAMCorrection12} while also accounting for multi-ion crosstalk \cite{Debnath16}.

The single-iteration, single-solution Grover search algorithm shown in Figure \ref{fig:choose1} has a theoretical ASP of 78.1\%, as discussed above. The SSO takes into account that the 7 unmarked states then have equal expected probabilies totaling 21.9\% of being measured. For all Boolean oracles, the average ASP is 38.9(4)\% and the average SSO is 83.2(7)\%, while phase oracles have an average ASP of 43.7(2)\% and an average SSO of 84.9(4)\%; the reduced use of resources in the phase oracles (10 $XX(\chi)$ gates and 3 qubits for phase oracles compared to 16 $XX(\chi)$ gates and 5 qubits for Boolean oracles) results in better performance, as expected. These results compare favorably with the classical ASP of 25\%.

The two-solution Grover search algorithm shown in Figure \ref{fig:choose2} has a theoretical ASP of 100\%, as discussed above. For all Boolean oracles, the average ASP is 67.9(2)\% and the average SSO is 67.6(2)\%, while phase oracles have an average ASP of 75.3(2)\% and an average SSO of 74.4(2)\%; the reduced use of resources in the phase oracles (6-8 $XX(\chi)$ gates and 3 qubits for phase oracles compared to 10-14 $XX(\chi)$ gates and 4 qubits for Boolean oracles) results in better performance, as expected. For all oracles in both cases, the two states with the highest measurement probability are also the two marked states. These results compare favorably with the classical ASP of 46.4\%.

We note that this implementation of the Grover search algorithm scales linearly in the two-qubit gate count and ancilla count for increasing search database size as a function of the number of qubits $n$, and for a constant number of solutions $t$. For a database of size $N=2^n$ stored on $n$ qubits, the amplification stage requires one Toffoli-$n$ gate, and the $t$-solution oracle stage requires at worst $t$ Toffoli-$n$ (for a phase oracle) or Toffoli-$(n+1)$ (for a Boolean oracle) gates; optimal oracles for particular sets of marked states may require even fewer two-qubit gates. 
The method used here to construct the Toffoli-4 circuit scales to Toffoli-$n$ gates as $6n-13$ in the two-qubit gate count and as $\lceil \frac{n-3}{2} \rceil$ in the ancilla count \cite{maslovToffoli16}. This paves the way for more extensive use of the Grover search algorithm in solving larger problems on quantum computers, including using the circuit as a subroutine for other quantum algorithms.

\begin{acknowledgments}
We thank S. Kimmel for helpful discussions. Circuits were drawn using the qcircuit.tex package. This work was supported by the ARO with funds from the IARPA LogiQ program, the AFOSR MURI program, and the NSF Physics Frontier Center at JQI. 
Declaration of competing financial interests: C.M. is a founding scientist of IonQ, Inc.

This material was partially based on work supported by the National Science Foundation during D.M.'s assignment at the Foundation. Any opinion, finding, and conclusions or recommendations expressed in this material are those of the authors and do not necessarily reflect the views of the National Science Foundation.
\end{acknowledgments}

\afterpage{\clearpage}


\bibliographystyle{apsrev4-1}

\begin{thebibliography}{30}%
\makeatletter
\providecommand \@ifxundefined [1]{%
 \@ifx{#1\undefined}
}%
\providecommand \@ifnum [1]{%
 \ifnum #1\expandafter \@firstoftwo
 \else \expandafter \@secondoftwo
 \fi
}%
\providecommand \@ifx [1]{%
 \ifx #1\expandafter \@firstoftwo
 \else \expandafter \@secondoftwo
 \fi
}%
\providecommand \natexlab [1]{#1}%
\providecommand \enquote  [1]{``#1''}%
\providecommand \bibnamefont  [1]{#1}%
\providecommand \bibfnamefont [1]{#1}%
\providecommand \citenamefont [1]{#1}%
\providecommand \href@noop [0]{\@secondoftwo}%
\providecommand \href [0]{\begingroup \@sanitize@url \@href}%
\providecommand \@href[1]{\@@startlink{#1}\@@href}%
\providecommand \@@href[1]{\endgroup#1\@@endlink}%
\providecommand \@sanitize@url [0]{\catcode `\\12\catcode `\$12\catcode
  `\&12\catcode `\#12\catcode `\^12\catcode `\_12\catcode `\%12\relax}%
\providecommand \@@startlink[1]{}%
\providecommand \@@endlink[0]{}%
\providecommand \url  [0]{\begingroup\@sanitize@url \@url }%
\providecommand \@url [1]{\endgroup\@href {#1}{\urlprefix }}%
\providecommand \urlprefix  [0]{URL }%
\providecommand \Eprint [0]{\href }%
\providecommand \doibase [0]{http://dx.doi.org/}%
\providecommand \selectlanguage [0]{\@gobble}%
\providecommand \bibinfo  [0]{\@secondoftwo}%
\providecommand \bibfield  [0]{\@secondoftwo}%
\providecommand \translation [1]{[#1]}%
\providecommand \BibitemOpen [0]{}%
\providecommand \bibitemStop [0]{}%
\providecommand \bibitemNoStop [0]{.\EOS\space}%
\providecommand \EOS [0]{\spacefactor3000\relax}%
\providecommand \BibitemShut  [1]{\csname bibitem#1\endcsname}%
\let\auto@bib@innerbib\@empty
\bibitem [{\citenamefont {Grover}(1997)}]{GroverOriginal97}%
  \BibitemOpen
  \bibfield  {author} {\bibinfo {author} {\bibfnamefont {L.~K.}\ \bibnamefont
  {Grover}},\ }\href {\doibase 10.1103/PhysRevLett.79.325} {\bibfield
  {journal} {\bibinfo  {journal} {Phys. Rev. Lett.}\ }\textbf {\bibinfo
  {volume} {79}},\ \bibinfo {pages} {325} (\bibinfo {year} {1997})}\BibitemShut
  {NoStop}%
\bibitem [{\citenamefont {Boyer}\ \emph {et~al.}(1998)\citenamefont {Boyer},
  \citenamefont {Brassard}, \citenamefont {H\o{}yer},\ and\ \citenamefont
  {Tapp}}]{BoyerGrover98}%
  \BibitemOpen
  \bibfield  {author} {\bibinfo {author} {\bibfnamefont {M.}~\bibnamefont
  {Boyer}}, \bibinfo {author} {\bibfnamefont {G.}~\bibnamefont {Brassard}},
  \bibinfo {author} {\bibfnamefont {P.}~\bibnamefont {H\o{}yer}}, \ and\
  \bibinfo {author} {\bibfnamefont {A.}~\bibnamefont {Tapp}},\ }\href {\doibase
  10.1002/(SICI)1521-3978(199806)46:4/5<493::AID-PROP493>3.0.CO;2-P} {\bibfield
   {journal} {\bibinfo  {journal} {Fortschr. Phys.}\ }\textbf {\bibinfo
  {volume} {46}},\ \bibinfo {pages} {493} (\bibinfo {year} {1998})}\BibitemShut
  {NoStop}%
\bibitem [{\citenamefont {Bennett}\ \emph {et~al.}(1997)\citenamefont
  {Bennett}, \citenamefont {Bernstein}, \citenamefont {Brassard},\ and\
  \citenamefont {Vazirani}}]{BennettBounds97}%
  \BibitemOpen
  \bibfield  {author} {\bibinfo {author} {\bibfnamefont {C.}~\bibnamefont
  {Bennett}}, \bibinfo {author} {\bibfnamefont {E.}~\bibnamefont {Bernstein}},
  \bibinfo {author} {\bibfnamefont {G.}~\bibnamefont {Brassard}}, \ and\
  \bibinfo {author} {\bibfnamefont {U.}~\bibnamefont {Vazirani}},\ }\href
  {\doibase 10.1137/S0097539796300933} {\bibfield  {journal} {\bibinfo
  {journal} {{SIAM} J. Comput.}\ }\textbf {\bibinfo {volume} {26}},\ \bibinfo
  {pages} {1510} (\bibinfo {year} {1997})}\BibitemShut {NoStop}%
\bibitem [{\citenamefont {Magniez}\ \emph {et~al.}(2007)\citenamefont
  {Magniez}, \citenamefont {Santha},\ and\ \citenamefont
  {Szegedy}}]{MagniezGroverSubroutine07}%
  \BibitemOpen
  \bibfield  {author} {\bibinfo {author} {\bibfnamefont {F.}~\bibnamefont
  {Magniez}}, \bibinfo {author} {\bibfnamefont {M.}~\bibnamefont {Santha}}, \
  and\ \bibinfo {author} {\bibfnamefont {M.}~\bibnamefont {Szegedy}},\ }\href
  {\doibase 10.1137/050643684} {\bibfield  {journal} {\bibinfo  {journal}
  {{SIAM} J. Comput.}\ }\textbf {\bibinfo {volume} {37}},\ \bibinfo {pages}
  {413} (\bibinfo {year} {2007})}\BibitemShut {NoStop}%
\bibitem [{\citenamefont {D{\"u}rr}\ \emph {et~al.}(2006)\citenamefont
  {D{\"u}rr}, \citenamefont {Heiligman}, \citenamefont {H\o{}yer},\ and\
  \citenamefont {Mhalla}}]{DurrGroverSubroutine06}%
  \BibitemOpen
  \bibfield  {author} {\bibinfo {author} {\bibfnamefont {C.}~\bibnamefont
  {D{\"u}rr}}, \bibinfo {author} {\bibfnamefont {M.}~\bibnamefont {Heiligman}},
  \bibinfo {author} {\bibfnamefont {P.}~\bibnamefont {H\o{}yer}}, \ and\
  \bibinfo {author} {\bibfnamefont {M.}~\bibnamefont {Mhalla}},\ }\href
  {\doibase 10.1137/050644719} {\bibfield  {journal} {\bibinfo  {journal}
  {{SIAM} J. Comput.}\ }\textbf {\bibinfo {volume} {35}},\ \bibinfo {pages}
  {1310} (\bibinfo {year} {2006})}\BibitemShut {NoStop}%
\bibitem [{\citenamefont {Chuang}\ \emph {et~al.}(1998)\citenamefont {Chuang},
  \citenamefont {Gershenfeld},\ and\ \citenamefont
  {Kubinec}}]{ChuangNMRGrover98}%
  \BibitemOpen
  \bibfield  {author} {\bibinfo {author} {\bibfnamefont {I.~L.}\ \bibnamefont
  {Chuang}}, \bibinfo {author} {\bibfnamefont {N.}~\bibnamefont {Gershenfeld}},
  \ and\ \bibinfo {author} {\bibfnamefont {M.}~\bibnamefont {Kubinec}},\ }\href
  {\doibase 10.1103/PhysRevLett.80.3408} {\bibfield  {journal} {\bibinfo
  {journal} {Phys. Rev. Lett.}\ }\textbf {\bibinfo {volume} {80}},\ \bibinfo
  {pages} {3408} (\bibinfo {year} {1998})}\BibitemShut {NoStop}%
\bibitem [{\citenamefont {Bhattacharya}\ \emph {et~al.}(2002)\citenamefont
  {Bhattacharya}, \citenamefont {van Linden van~den Heuvell},\ and\
  \citenamefont {Spreeuw}}]{BhattacharyaFourierOptics02}%
  \BibitemOpen
  \bibfield  {author} {\bibinfo {author} {\bibfnamefont {N.}~\bibnamefont
  {Bhattacharya}}, \bibinfo {author} {\bibfnamefont {H.~B.}\ \bibnamefont {van
  Linden van~den Heuvell}}, \ and\ \bibinfo {author} {\bibfnamefont {R.~J.~C.}\
  \bibnamefont {Spreeuw}},\ }\href {\doibase 10.1103/PhysRevLett.88.137901}
  {\bibfield  {journal} {\bibinfo  {journal} {Phys. Rev. Lett.}\ }\textbf
  {\bibinfo {volume} {88}},\ \bibinfo {pages} {137901} (\bibinfo {year}
  {2002})}\BibitemShut {NoStop}%
\bibitem [{\citenamefont {Brickman}\ \emph {et~al.}(2005)\citenamefont
  {Brickman}, \citenamefont {Haljan}, \citenamefont {Lee}, \citenamefont
  {Acton}, \citenamefont {Deslauriers},\ and\ \citenamefont
  {Monroe}}]{BrickmanIons05}%
  \BibitemOpen
  \bibfield  {author} {\bibinfo {author} {\bibfnamefont {K.-A.}\ \bibnamefont
  {Brickman}}, \bibinfo {author} {\bibfnamefont {P.~C.}\ \bibnamefont
  {Haljan}}, \bibinfo {author} {\bibfnamefont {P.~J.}\ \bibnamefont {Lee}},
  \bibinfo {author} {\bibfnamefont {M.}~\bibnamefont {Acton}}, \bibinfo
  {author} {\bibfnamefont {L.}~\bibnamefont {Deslauriers}}, \ and\ \bibinfo
  {author} {\bibfnamefont {C.}~\bibnamefont {Monroe}},\ }\href
  {http://link.aps.org/doi/10.1103/PhysRevA.72.050306} {\bibfield  {journal}
  {\bibinfo  {journal} {Physical Review A}\ }\textbf {\bibinfo {volume} {72}},\
  \bibinfo {pages} {050306(R)} (\bibinfo {year} {2005})}\BibitemShut {NoStop}%
\bibitem [{\citenamefont {Walther}\ \emph {et~al.}(2005)\citenamefont
  {Walther}, \citenamefont {Resch}, \citenamefont {Rudolph}, \citenamefont
  {Schenck}, \citenamefont {Weinfurter}, \citenamefont {Vedral}, \citenamefont
  {Aspelmeyer},\ and\ \citenamefont {Zeilinger}}]{WaltherPhotonClusters05}%
  \BibitemOpen
  \bibfield  {author} {\bibinfo {author} {\bibfnamefont {P.}~\bibnamefont
  {Walther}}, \bibinfo {author} {\bibfnamefont {K.~J.}\ \bibnamefont {Resch}},
  \bibinfo {author} {\bibfnamefont {T.}~\bibnamefont {Rudolph}}, \bibinfo
  {author} {\bibfnamefont {E.}~\bibnamefont {Schenck}}, \bibinfo {author}
  {\bibfnamefont {H.}~\bibnamefont {Weinfurter}}, \bibinfo {author}
  {\bibfnamefont {V.}~\bibnamefont {Vedral}}, \bibinfo {author} {\bibfnamefont
  {M.}~\bibnamefont {Aspelmeyer}}, \ and\ \bibinfo {author} {\bibfnamefont
  {A.}~\bibnamefont {Zeilinger}},\ }\href {\doibase 10.1038/nature03347}
  {\bibfield  {journal} {\bibinfo  {journal} {Nature}\ }\textbf {\bibinfo
  {volume} {434}},\ \bibinfo {pages} {169} (\bibinfo {year}
  {2005})}\BibitemShut {NoStop}%
\bibitem [{\citenamefont {DiCarlo}\ \emph {et~al.}(2009)\citenamefont
  {DiCarlo}, \citenamefont {Chow}, \citenamefont {Gambetta}, \citenamefont
  {Bishop}, \citenamefont {Johnson}, \citenamefont {Schuster}, \citenamefont
  {Majer}, \citenamefont {Blais}, \citenamefont {Frunzio}, \citenamefont
  {Girvin},\ and\ \citenamefont {Schoelkopf}}]{DicarloSC09}%
  \BibitemOpen
  \bibfield  {author} {\bibinfo {author} {\bibfnamefont {L.}~\bibnamefont
  {DiCarlo}}, \bibinfo {author} {\bibfnamefont {J.~M.}\ \bibnamefont {Chow}},
  \bibinfo {author} {\bibfnamefont {J.~M.}\ \bibnamefont {Gambetta}}, \bibinfo
  {author} {\bibfnamefont {L.~S.}\ \bibnamefont {Bishop}}, \bibinfo {author}
  {\bibfnamefont {B.~R.}\ \bibnamefont {Johnson}}, \bibinfo {author}
  {\bibfnamefont {D.~I.}\ \bibnamefont {Schuster}}, \bibinfo {author}
  {\bibfnamefont {J.}~\bibnamefont {Majer}}, \bibinfo {author} {\bibfnamefont
  {A.}~\bibnamefont {Blais}}, \bibinfo {author} {\bibfnamefont
  {L.}~\bibnamefont {Frunzio}}, \bibinfo {author} {\bibfnamefont {S.~M.}\
  \bibnamefont {Girvin}}, \ and\ \bibinfo {author} {\bibfnamefont {R.~J.}\
  \bibnamefont {Schoelkopf}},\ }\href {\doibase 10.1038/nature08121} {\bibfield
   {journal} {\bibinfo  {journal} {Nature}\ }\textbf {\bibinfo {volume}
  {460}},\ \bibinfo {pages} {240} (\bibinfo {year} {2009})}\BibitemShut
  {NoStop}%
\bibitem [{\citenamefont {Barz}\ \emph {et~al.}(2012)\citenamefont {Barz},
  \citenamefont {Kashefi}, \citenamefont {Broadbent}, \citenamefont
  {Fitzsimons}, \citenamefont {Zeilinger},\ and\ \citenamefont
  {Walther}}]{BarzPhotonics12}%
  \BibitemOpen
  \bibfield  {author} {\bibinfo {author} {\bibfnamefont {S.}~\bibnamefont
  {Barz}}, \bibinfo {author} {\bibfnamefont {E.}~\bibnamefont {Kashefi}},
  \bibinfo {author} {\bibfnamefont {A.}~\bibnamefont {Broadbent}}, \bibinfo
  {author} {\bibfnamefont {J.~F.}\ \bibnamefont {Fitzsimons}}, \bibinfo
  {author} {\bibfnamefont {A.}~\bibnamefont {Zeilinger}}, \ and\ \bibinfo
  {author} {\bibfnamefont {P.}~\bibnamefont {Walther}},\ }\href {\doibase
  10.1126/science.1214707} {\bibfield  {journal} {\bibinfo  {journal}
  {Science}\ }\textbf {\bibinfo {volume} {335}},\ \bibinfo {pages} {303}
  (\bibinfo {year} {2012})}\BibitemShut {NoStop}%
\bibitem [{\citenamefont {M\o{}lmer}\ \emph {et~al.}(2011)\citenamefont
  {M\o{}lmer}, \citenamefont {Isenhower},\ and\ \citenamefont
  {Saffman}}]{MolmerRydProposal11}%
  \BibitemOpen
  \bibfield  {author} {\bibinfo {author} {\bibfnamefont {K.}~\bibnamefont
  {M\o{}lmer}}, \bibinfo {author} {\bibfnamefont {L.}~\bibnamefont
  {Isenhower}}, \ and\ \bibinfo {author} {\bibfnamefont {M.}~\bibnamefont
  {Saffman}},\ }\href {\doibase 10.1088/0953-4075/44/18/184016} {\bibfield
  {journal} {\bibinfo  {journal} {J. Phys. B: At. Mol. Opt. Phys.}\ }\textbf
  {\bibinfo {volume} {44}},\ \bibinfo {pages} {184016} (\bibinfo {year}
  {2011})}\BibitemShut {NoStop}%
\bibitem [{\citenamefont {Vandersypen}\ \emph {et~al.}(2000)\citenamefont
  {Vandersypen}, \citenamefont {Steffen}, \citenamefont {Sherwood},
  \citenamefont {Yannoni}, \citenamefont {Breyta},\ and\ \citenamefont
  {Chuang}}]{Vandersypen3qubitNMR00}%
  \BibitemOpen
  \bibfield  {author} {\bibinfo {author} {\bibfnamefont {L.~M.~K.}\
  \bibnamefont {Vandersypen}}, \bibinfo {author} {\bibfnamefont
  {M.}~\bibnamefont {Steffen}}, \bibinfo {author} {\bibfnamefont {M.~H.}\
  \bibnamefont {Sherwood}}, \bibinfo {author} {\bibfnamefont {C.~S.}\
  \bibnamefont {Yannoni}}, \bibinfo {author} {\bibfnamefont {G.}~\bibnamefont
  {Breyta}}, \ and\ \bibinfo {author} {\bibfnamefont {I.~L.}\ \bibnamefont
  {Chuang}},\ }\href {\doibase 10.1063/1.125846} {\bibfield  {journal}
  {\bibinfo  {journal} {Appl. Phys. Lett.}\ }\textbf {\bibinfo {volume} {76}},\
  \bibinfo {pages} {646} (\bibinfo {year} {2000})}\BibitemShut {NoStop}%
\bibitem [{\citenamefont {Debnath}\ \emph {et~al.}(2016)\citenamefont
  {Debnath}, \citenamefont {Linke}, \citenamefont {Figgatt}, \citenamefont
  {Landsman}, \citenamefont {Wright},\ and\ \citenamefont
  {Monroe}}]{Debnath16}%
  \BibitemOpen
  \bibfield  {author} {\bibinfo {author} {\bibfnamefont {S.}~\bibnamefont
  {Debnath}}, \bibinfo {author} {\bibfnamefont {N.~M.}\ \bibnamefont {Linke}},
  \bibinfo {author} {\bibfnamefont {C.}~\bibnamefont {Figgatt}}, \bibinfo
  {author} {\bibfnamefont {K.~A.}\ \bibnamefont {Landsman}}, \bibinfo {author}
  {\bibfnamefont {K.}~\bibnamefont {Wright}}, \ and\ \bibinfo {author}
  {\bibfnamefont {C.}~\bibnamefont {Monroe}},\ }\href {\doibase
  10.1038/nature18648} {\bibfield  {journal} {\bibinfo  {journal} {Nature}\
  }\textbf {\bibinfo {volume} {536}},\ \bibinfo {pages} {63} (\bibinfo {year}
  {2016})}\BibitemShut {NoStop}%
\bibitem [{\citenamefont {Nielsen}\ and\ \citenamefont
  {Chuang}(2011)}]{NielsenChuang11}%
  \BibitemOpen
  \bibfield  {author} {\bibinfo {author} {\bibfnamefont {M.~A.}\ \bibnamefont
  {Nielsen}}\ and\ \bibinfo {author} {\bibfnamefont {I.~L.}\ \bibnamefont
  {Chuang}},\ }\href@noop {} {\emph {\bibinfo {title} {Quantum Computation and
  Quantum Information: 10th Anniversary Edition}}},\ \bibinfo {edition} {10th}\
  ed.\ (\bibinfo  {publisher} {Cambridge University Press},\ \bibinfo {address}
  {New York, NY, USA},\ \bibinfo {year} {2011})\BibitemShut {NoStop}%
\bibitem [{\citenamefont {Milburn}\ \emph {et~al.}(2000)\citenamefont
  {Milburn}, \citenamefont {Schneider},\ and\ \citenamefont
  {James}}]{Milburn00}%
  \BibitemOpen
  \bibfield  {author} {\bibinfo {author} {\bibfnamefont {G.}~\bibnamefont
  {Milburn}}, \bibinfo {author} {\bibfnamefont {S.}~\bibnamefont {Schneider}},
  \ and\ \bibinfo {author} {\bibfnamefont {D.}~\bibnamefont {James}},\ }\href
  {\doibase 10.1002/1521-3978(200009)48:9/11<801::AID-PROP801>3.0.CO;2-1}
  {\bibfield  {journal} {\bibinfo  {journal} {Fortschr. Phys.}\ }\textbf
  {\bibinfo {volume} {48}},\ \bibinfo {pages} {801} (\bibinfo {year}
  {2000})}\BibitemShut {NoStop}%
\bibitem [{\citenamefont {Olmschenk}\ \emph {et~al.}(2007)\citenamefont
  {Olmschenk}, \citenamefont {Younge}, \citenamefont {Moehring}, \citenamefont
  {Matsukevich}, \citenamefont {Maunz},\ and\ \citenamefont
  {Monroe}}]{Olmschenk07}%
  \BibitemOpen
  \bibfield  {author} {\bibinfo {author} {\bibfnamefont {S.}~\bibnamefont
  {Olmschenk}}, \bibinfo {author} {\bibfnamefont {K.~C.}\ \bibnamefont
  {Younge}}, \bibinfo {author} {\bibfnamefont {D.~L.}\ \bibnamefont
  {Moehring}}, \bibinfo {author} {\bibfnamefont {D.~N.}\ \bibnamefont
  {Matsukevich}}, \bibinfo {author} {\bibfnamefont {P.}~\bibnamefont {Maunz}},
  \ and\ \bibinfo {author} {\bibfnamefont {C.}~\bibnamefont {Monroe}},\ }\href
  {\doibase 10.1103/PhysRevA.76.052314} {\bibfield  {journal} {\bibinfo
  {journal} {Phys. Rev. A}\ }\textbf {\bibinfo {volume} {76}},\ \bibinfo
  {pages} {052314} (\bibinfo {year} {2007})}\BibitemShut {NoStop}%
\bibitem [{\citenamefont {Hayes}\ \emph {et~al.}(2010)\citenamefont {Hayes},
  \citenamefont {Matsukevich}, \citenamefont {Maunz}, \citenamefont {Hucul},
  \citenamefont {Quraishi}, \citenamefont {Olmschenk}, \citenamefont
  {Campbell}, \citenamefont {Mizrahi}, \citenamefont {Senko},\ and\
  \citenamefont {Monroe}}]{HayesCombs10}%
  \BibitemOpen
  \bibfield  {author} {\bibinfo {author} {\bibfnamefont {D.}~\bibnamefont
  {Hayes}}, \bibinfo {author} {\bibfnamefont {D.~N.}\ \bibnamefont
  {Matsukevich}}, \bibinfo {author} {\bibfnamefont {P.}~\bibnamefont {Maunz}},
  \bibinfo {author} {\bibfnamefont {D.}~\bibnamefont {Hucul}}, \bibinfo
  {author} {\bibfnamefont {Q.}~\bibnamefont {Quraishi}}, \bibinfo {author}
  {\bibfnamefont {S.}~\bibnamefont {Olmschenk}}, \bibinfo {author}
  {\bibfnamefont {W.}~\bibnamefont {Campbell}}, \bibinfo {author}
  {\bibfnamefont {J.}~\bibnamefont {Mizrahi}}, \bibinfo {author} {\bibfnamefont
  {C.}~\bibnamefont {Senko}}, \ and\ \bibinfo {author} {\bibfnamefont
  {C.}~\bibnamefont {Monroe}},\ }\href {\doibase
  10.1103/PhysRevLett.104.140501} {\bibfield  {journal} {\bibinfo  {journal}
  {Phys. Rev. Lett.}\ }\textbf {\bibinfo {volume} {104}},\ \bibinfo {pages}
  {140501} (\bibinfo {year} {2010})}\BibitemShut {NoStop}%
\bibitem [{\citenamefont {Solano}\ \emph {et~al.}(1999)\citenamefont {Solano},
  \citenamefont {de~Matos~Filho},\ and\ \citenamefont {Zagury}}]{Solano99}%
  \BibitemOpen
  \bibfield  {author} {\bibinfo {author} {\bibfnamefont {E.}~\bibnamefont
  {Solano}}, \bibinfo {author} {\bibfnamefont {R.~L.}\ \bibnamefont
  {de~Matos~Filho}}, \ and\ \bibinfo {author} {\bibfnamefont {N.}~\bibnamefont
  {Zagury}},\ }\href {\doibase 10.1103/PhysRevA.59.R2539} {\bibfield  {journal}
  {\bibinfo  {journal} {Phys. Rev. A}\ }\textbf {\bibinfo {volume} {59}},\
  \bibinfo {pages} {R2539} (\bibinfo {year} {1999})}\BibitemShut {NoStop}%
\bibitem [{\citenamefont {M\o{}lmer}\ and\ \citenamefont
  {S\o{}rensen}(1999)}]{Molmer99}%
  \BibitemOpen
  \bibfield  {author} {\bibinfo {author} {\bibfnamefont {K.}~\bibnamefont
  {M\o{}lmer}}\ and\ \bibinfo {author} {\bibfnamefont {A.}~\bibnamefont
  {S\o{}rensen}},\ }\href {\doibase 10.1103/PhysRevLett.82.1835} {\bibfield
  {journal} {\bibinfo  {journal} {Phys. Rev. Lett.}\ }\textbf {\bibinfo
  {volume} {82}},\ \bibinfo {pages} {1835} (\bibinfo {year}
  {1999})}\BibitemShut {NoStop}%
\bibitem [{\citenamefont {Zhu}\ \emph {et~al.}(2006)\citenamefont {Zhu},
  \citenamefont {Monroe},\ and\ \citenamefont {Duan}}]{ZhuGates06}%
  \BibitemOpen
  \bibfield  {author} {\bibinfo {author} {\bibfnamefont {S.-L.}\ \bibnamefont
  {Zhu}}, \bibinfo {author} {\bibfnamefont {C.}~\bibnamefont {Monroe}}, \ and\
  \bibinfo {author} {\bibfnamefont {L.-M.}\ \bibnamefont {Duan}},\ }\href
  {\doibase 10.1209/epl/i2005-10424-4} {\bibfield  {journal} {\bibinfo
  {journal} {Europhys. Lett.}\ }\textbf {\bibinfo {volume} {73}},\ \bibinfo
  {pages} {485} (\bibinfo {year} {2006})}\BibitemShut {NoStop}%
\bibitem [{\citenamefont {Choi}\ \emph {et~al.}(2014)\citenamefont {Choi},
  \citenamefont {Debnath}, \citenamefont {Manning}, \citenamefont {Figgatt},
  \citenamefont {Gong}, \citenamefont {Duan},\ and\ \citenamefont
  {Monroe}}]{Choi14}%
  \BibitemOpen
  \bibfield  {author} {\bibinfo {author} {\bibfnamefont {T.}~\bibnamefont
  {Choi}}, \bibinfo {author} {\bibfnamefont {S.}~\bibnamefont {Debnath}},
  \bibinfo {author} {\bibfnamefont {T.~A.}\ \bibnamefont {Manning}}, \bibinfo
  {author} {\bibfnamefont {C.}~\bibnamefont {Figgatt}}, \bibinfo {author}
  {\bibfnamefont {Z.-X.}\ \bibnamefont {Gong}}, \bibinfo {author}
  {\bibfnamefont {L.-M.}\ \bibnamefont {Duan}}, \ and\ \bibinfo {author}
  {\bibfnamefont {C.}~\bibnamefont {Monroe}},\ }\href {\doibase
  10.1103/PhysRevLett.112.190502} {\bibfield  {journal} {\bibinfo  {journal}
  {Phys. Rev. Lett.}\ }\textbf {\bibinfo {volume} {112}},\ \bibinfo {pages}
  {190502} (\bibinfo {year} {2014})}\BibitemShut {NoStop}%
\bibitem [{\citenamefont {Cory}\ \emph {et~al.}(1998)\citenamefont {Cory},
  \citenamefont {Price},\ and\ \citenamefont {Havel}}]{CoryNMRToffoli98}%
  \BibitemOpen
  \bibfield  {author} {\bibinfo {author} {\bibfnamefont {D.~G.}\ \bibnamefont
  {Cory}}, \bibinfo {author} {\bibfnamefont {M.~D.}\ \bibnamefont {Price}}, \
  and\ \bibinfo {author} {\bibfnamefont {T.~F.}\ \bibnamefont {Havel}},\ }\href
  {\doibase 10.1016/S0167-2789(98)00046-3} {\bibfield  {journal} {\bibinfo
  {journal} {Physica D}\ }\textbf {\bibinfo {volume} {120}},\ \bibinfo {pages}
  {82} (\bibinfo {year} {1998})}\BibitemShut {NoStop}%
\bibitem [{\citenamefont {Monz}\ \emph {et~al.}(2009)\citenamefont {Monz},
  \citenamefont {Kim}, \citenamefont {H\"ansel}, \citenamefont {Riebe},
  \citenamefont {Villar}, \citenamefont {Schindler}, \citenamefont {Chwalla},
  \citenamefont {Hennrich},\ and\ \citenamefont {Blatt}}]{Monz09}%
  \BibitemOpen
  \bibfield  {author} {\bibinfo {author} {\bibfnamefont {T.}~\bibnamefont
  {Monz}}, \bibinfo {author} {\bibfnamefont {K.}~\bibnamefont {Kim}}, \bibinfo
  {author} {\bibfnamefont {W.}~\bibnamefont {H\"ansel}}, \bibinfo {author}
  {\bibfnamefont {M.}~\bibnamefont {Riebe}}, \bibinfo {author} {\bibfnamefont
  {A.~S.}\ \bibnamefont {Villar}}, \bibinfo {author} {\bibfnamefont
  {P.}~\bibnamefont {Schindler}}, \bibinfo {author} {\bibfnamefont
  {M.}~\bibnamefont {Chwalla}}, \bibinfo {author} {\bibfnamefont
  {M.}~\bibnamefont {Hennrich}}, \ and\ \bibinfo {author} {\bibfnamefont
  {R.}~\bibnamefont {Blatt}},\ }\href {\doibase 10.1103/PhysRevLett.102.040501}
  {\bibfield  {journal} {\bibinfo  {journal} {Phys. Rev. Lett.}\ }\textbf
  {\bibinfo {volume} {102}},\ \bibinfo {pages} {040501} (\bibinfo {year}
  {2009})}\BibitemShut {NoStop}%
\bibitem [{\citenamefont {Linke}\ \emph {et~al.}(2017)\citenamefont {Linke},
  \citenamefont {Maslov}, \citenamefont {Roetteler}, \citenamefont {Debnath},
  \citenamefont {Figgatt}, \citenamefont {Landsman}, \citenamefont {Wright},\
  and\ \citenamefont {Monroe}}]{LinkeComparison17}%
  \BibitemOpen
  \bibfield  {author} {\bibinfo {author} {\bibfnamefont {N.~M.}\ \bibnamefont
  {Linke}}, \bibinfo {author} {\bibfnamefont {D.}~\bibnamefont {Maslov}},
  \bibinfo {author} {\bibfnamefont {M.}~\bibnamefont {Roetteler}}, \bibinfo
  {author} {\bibfnamefont {S.}~\bibnamefont {Debnath}}, \bibinfo {author}
  {\bibfnamefont {C.}~\bibnamefont {Figgatt}}, \bibinfo {author} {\bibfnamefont
  {K.~A.}\ \bibnamefont {Landsman}}, \bibinfo {author} {\bibfnamefont
  {K.}~\bibnamefont {Wright}}, \ and\ \bibinfo {author} {\bibfnamefont
  {C.}~\bibnamefont {Monroe}},\ }\href {http://arxiv.org/abs/1702.01852}
  {\bibfield  {journal} {\bibinfo  {journal} {PNAS}\ }\textbf {\bibinfo
  {volume} {114}},\ \bibinfo {pages} {3305} (\bibinfo
  {year} {2017})}\BibitemShut {NoStop}%
\bibitem [{\citenamefont {Barenco}\ \emph {et~al.}(1995)\citenamefont
  {Barenco}, \citenamefont {Bennett}, \citenamefont {Cleve}, \citenamefont
  {DiVincenzo}, \citenamefont {Margolus}, \citenamefont {Shor}, \citenamefont
  {Sleator}, \citenamefont {Smolin},\ and\ \citenamefont
  {Weinfurter}}]{Barenco95}%
  \BibitemOpen
  \bibfield  {author} {\bibinfo {author} {\bibfnamefont {A.}~\bibnamefont
  {Barenco}}, \bibinfo {author} {\bibfnamefont {C.~H.}\ \bibnamefont
  {Bennett}}, \bibinfo {author} {\bibfnamefont {R.}~\bibnamefont {Cleve}},
  \bibinfo {author} {\bibfnamefont {D.~P.}\ \bibnamefont {DiVincenzo}},
  \bibinfo {author} {\bibfnamefont {N.}~\bibnamefont {Margolus}}, \bibinfo
  {author} {\bibfnamefont {P.}~\bibnamefont {Shor}}, \bibinfo {author}
  {\bibfnamefont {T.}~\bibnamefont {Sleator}}, \bibinfo {author} {\bibfnamefont
  {J.~A.}\ \bibnamefont {Smolin}}, \ and\ \bibinfo {author} {\bibfnamefont
  {H.}~\bibnamefont {Weinfurter}},\ }\href {\doibase 10.1103/PhysRevA.52.3457}
  {\bibfield  {journal} {\bibinfo  {journal} {Phys. Rev. A}\ }\textbf {\bibinfo
  {volume} {52}},\ \bibinfo {pages} {3457} (\bibinfo {year}
  {1995})}\BibitemShut {NoStop}%
\bibitem [{\citenamefont {Maslov}(2017)}]{MaslovCircuits17}%
  \BibitemOpen
  \bibfield  {author} {\bibinfo {author} {\bibfnamefont {D.}~\bibnamefont
  {Maslov}},\ }\href {\doibase 10.1088/1367-2630/aa5e47} {\bibfield  {journal}
  {\bibinfo  {journal} {New J. Phys.}\ }\textbf {\bibinfo {volume} {19}},\
  \bibinfo {pages} {023035} (\bibinfo {year} {2017})}\BibitemShut {NoStop}%
\bibitem [{\citenamefont {Maslov}(2016)}]{maslovToffoli16}%
  \BibitemOpen
  \bibfield  {author} {\bibinfo {author} {\bibfnamefont {D.}~\bibnamefont
  {Maslov}},\ }\href {\doibase 10.1103/PhysRevA.93.022311} {\bibfield
  {journal} {\bibinfo  {journal} {Phys. Rev. A}\ }\textbf {\bibinfo {volume}
  {93}},\ \bibinfo {pages} {022311} (\bibinfo {year} {2016})}\BibitemShut
  {NoStop}%
\bibitem [{\citenamefont {Chiaverini}\ \emph {et~al.}(2005)\citenamefont
  {Chiaverini}, \citenamefont {Britton}, \citenamefont {Leibfried},
  \citenamefont {Knill}, \citenamefont {Barrett}, \citenamefont {Blakestad},
  \citenamefont {Itano}, \citenamefont {Jost}, \citenamefont {Langer},
  \citenamefont {Ozeri}, \citenamefont {Schaetz},\ and\ \citenamefont
  {Wineland}}]{ChiaveriniSSO05}%
  \BibitemOpen
  \bibfield  {author} {\bibinfo {author} {\bibfnamefont {J.}~\bibnamefont
  {Chiaverini}}, \bibinfo {author} {\bibfnamefont {J.}~\bibnamefont {Britton}},
  \bibinfo {author} {\bibfnamefont {D.}~\bibnamefont {Leibfried}}, \bibinfo
  {author} {\bibfnamefont {E.}~\bibnamefont {Knill}}, \bibinfo {author}
  {\bibfnamefont {M.~D.}\ \bibnamefont {Barrett}}, \bibinfo {author}
  {\bibfnamefont {R.~B.}\ \bibnamefont {Blakestad}}, \bibinfo {author}
  {\bibfnamefont {W.~M.}\ \bibnamefont {Itano}}, \bibinfo {author}
  {\bibfnamefont {J.~D.}\ \bibnamefont {Jost}}, \bibinfo {author}
  {\bibfnamefont {C.}~\bibnamefont {Langer}}, \bibinfo {author} {\bibfnamefont
  {R.}~\bibnamefont {Ozeri}}, \bibinfo {author} {\bibfnamefont
  {T.}~\bibnamefont {Schaetz}}, \ and\ \bibinfo {author} {\bibfnamefont
  {D.~J.}\ \bibnamefont {Wineland}},\ }\href {\doibase 10.1126/science.1110335}
  {\bibfield  {journal} {\bibinfo  {journal} {Science}\ }\textbf {\bibinfo
  {volume} {308}},\ \bibinfo {pages} {997} (\bibinfo {year}
  {2005})}\BibitemShut {NoStop}%
\bibitem [{\citenamefont {Shen}\ and\ \citenamefont
  {Duan}(2012)}]{ShenSPAMCorrection12}%
  \BibitemOpen
  \bibfield  {author} {\bibinfo {author} {\bibfnamefont {C.}~\bibnamefont
  {Shen}}\ and\ \bibinfo {author} {\bibfnamefont {L.-M.}\ \bibnamefont
  {Duan}},\ }\href {\doibase 10.1088/1367-2630/14/5/053053} {\bibfield
  {journal} {\bibinfo  {journal} {New J. Phys.}\ }\textbf {\bibinfo {volume}
  {14}},\ \bibinfo {pages} {053053} (\bibinfo {year} {2012})}\BibitemShut
  {NoStop}%
\end{thebibliography}
%

\newpage
\newpage
\pagebreak

\appendix*

\newpage
\pagebreak

\onecolumngrid
\section{\normalsize Supplementary Materials}

\textbf{Circuit Diagrams:} Here we present detailed circuit diagrams for all of the operations presented in the paper above, shown in terms of the $R(\theta, \phi)$ and $XX(\chi)$ gates directly implemented by the experiment. The single-qubit rotation is defined as
\begin{equation*}
R(\theta, \phi) = 
\begin{pmatrix}
\cos\frac{\theta}{2} & -ie^{-i\phi}\sin{\frac{\theta}{2}}\\[0.3em]
-ie^{i\phi}\sin\frac{\theta}{2} & \cos\frac{\theta}{2}
\end{pmatrix}.
\end{equation*}
Rotations about the $X$-axis ($R_x (\theta)$) are achieved by setting $\phi=0$, and rotations about the $Y$-axis ($R_y (\theta)$) are achieved by setting $\phi=\frac{\pi}{2}$. Rotations about the $Z$ axis ($R_z (\theta)$) are comprised of three rotations about axes in the $XY$ plane, as demonstrated in Figure \ref{fig:ZCircuit}.

\ZCircuit

The two-qubit entangling gate is 
\begin{equation*}
XX(\chi)=\begin{pmatrix}
\cos(\chi) & 0 & 0 & \hspace{-2mm}-i \sin(\chi)\\[0.3em]
0 & \hspace{-1mm}\cos(\chi) & \hspace{-3mm}-i \sin(\chi) & 0\\[0.3em]
0 & \hspace{-3mm}-i \sin(\chi) & \hspace{-1mm}\cos(\chi) & 0\\[0.3em]
-i \sin(\chi) & 0 & 0 & \cos(\chi)
\end{pmatrix}.
\end{equation*}
The parameter $\chi$ can be varied continuously by adjusting the overall power applied to the gate, but the gates used here require only $\chi=\pm \frac{\pi}{4}$ or $\chi=\pm \frac{\pi}{8}$. The gate is maximally entangling for $\chi=\pm\frac{\pi}{4}$, so $XX \left(\frac{\pi}{4}\right) |00\rangle = \frac{1}{\sqrt{2}} \left( |00\rangle - i |11\rangle\right)$.

Two-qubit $XX$ gates are combined with rotation $R$ gates to construct the composite gates needed for the Grover search algorithm implementation. The parameter $\chi$ can be positive or negative, depending on what ion pair is chosen and the particulars of the pulse segmentation solution chosen for the ion pair in question; the sign of $\chi$ ($\text{sgn}(\chi)$) is determined experimentally for each ion pair. Consequently, some composite gate circuits include rotations with parameters that depend on $\text{sgn}(\chi)$.

\TwoQubitCircuits

The two-qubit controlled-$NOT$ and controlled-$Z$ gates are shown in Figures \ref{fig:2QCircuits}(a-b). They each require one $XX$ gate and several rotations. The three-qubit gates used here are the Toffoli-3 and controlled-controlled-$Z$ ($CCZ$) gates, shown in Figures \ref{fig:3QCircuits}(a-b). The Toffoli-3 gate requires two control qubits ($q_1$ and $q_2$) and one target qubit ($q_t$). Finally, the four-qubit Toffoli-4 gate is shown in Figure \ref{fig:Toff4Circuit}. It governs a four-qubit interaction between 3 control qubits  ($q_1$, $q_2$, and $q_3$) and one target qubit ($q_t$), and it additionally requires an ancilla qubit ($q_a$).

\ThreeQubitCircuits

\ToffFourCircuit

The Grover search algorithm is implemented using circuits that are equivalent to those shown in Figures \ref{fig:GroverConcept}(b,d), but with the initialization and amplification stages optimized to minimize gate times, as shown in Figures \ref{fig:GroverCircuits}(a-b). The circuits shown are for use with Boolean oracles; in the phase oracle case, the ancilla qubit $q_a$ is simply omitted. To preserve the modularity of the algorithm, the initialization stage and amplification stage were each optimized without regard to the contents of the oracle, so each possible oracle can simply be inserted into the algorithm without making any changes to the other stages. The oracles were implemented as per the circuit diagrams shown in Table \ref{tab:Grover1Oracles} for single-solution oracles and Table \ref{tab:Grover2Oracles} for two-solution oracles.

\GroverCircuits

\begin{figure}
\begin{tabular}[c]{c c c c c c c}
\multicolumn{1}{l}{\bf{(a) Toffoli-3 Characterization Circuit}} & & & & & & \multicolumn{1}{l}{\bf{(b) Toffoli-3 Characterization}}\\

\multicolumn{1}{c}{\ToffTestCircuit} 
& & & & & &
\includegraphics[width=0.49\columnwidth,valign=T]{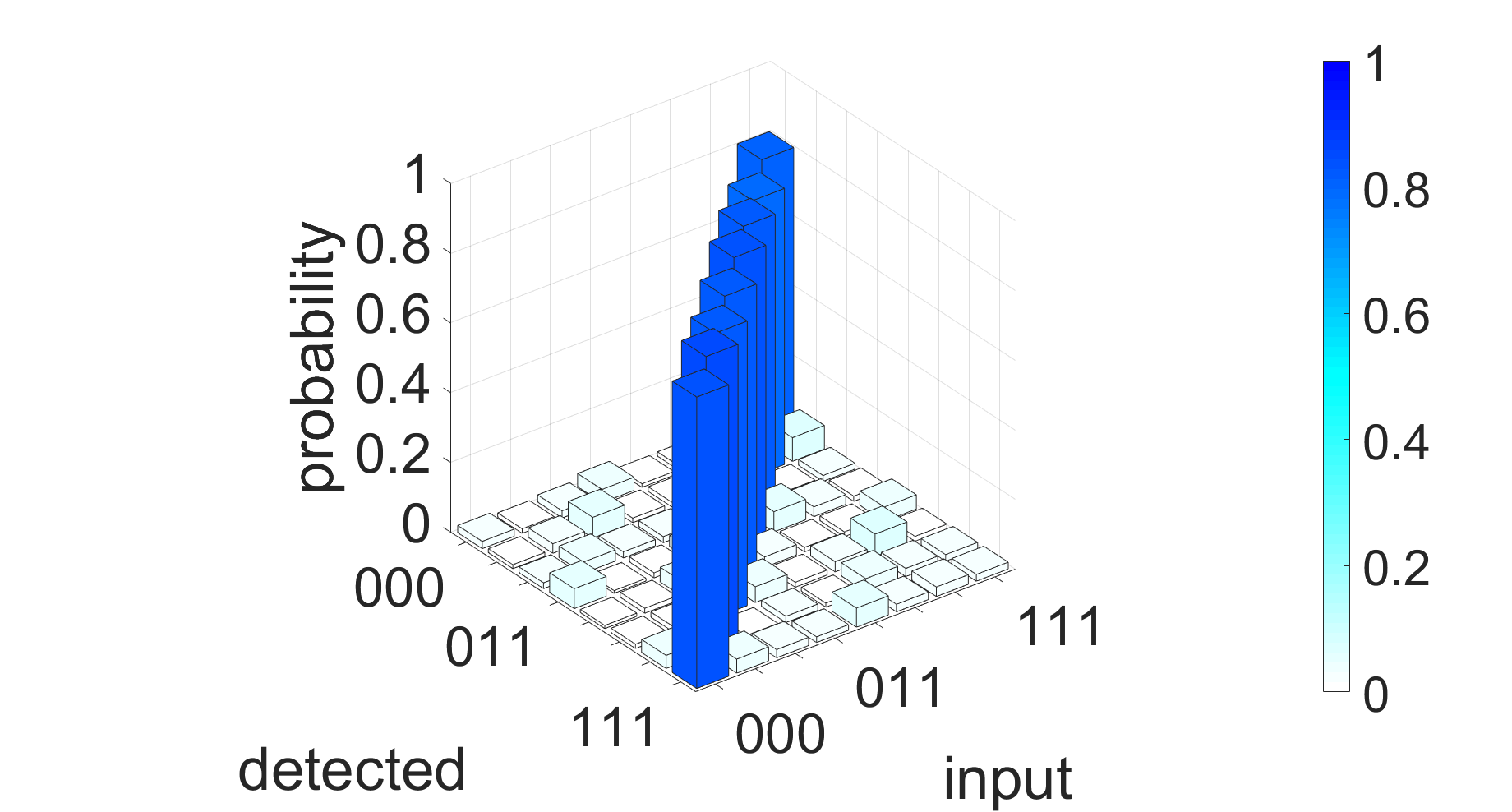}\\
\end{tabular}
\caption{(a) Circuit for implementing the Toffoli-3 limited tomography procedure. The global rotations are positive for even input states and negative for odd input states. (b) Limited tomography check performed on the Toffoli-3 gate to verify phases. The average success probability is 82.1(2)\%, corrected for a 2.4\% average SPAM error.}
\label{fig:ToffoliCheck}
\end{figure}

\textbf{Toffoli-3 Characterization:} We employed a limited tomography procedure to characterize the outputs of the Toffoli-3 gate performed. A global rotation into the $X$ basis was applied to all 3 ions before and after the Toffoli-3 gate for each input (see Figure \ref{fig:ToffoliCheck}(a)): $R_y(\frac{\pi}{2})$ for the even inputs ($000$, $010$, $100$, $110$) and $R_y(-\frac{\pi}{2})$ for the odd inputs ($001$, $011$, $101$, $111$). An ideal Toffoli-3 gate will result in an anti-diagonal input-output matrix in the $Z$ basis when this procedure is applied. The experimental results of this verification procedure are shown in Figure \ref{fig:ToffoliCheck}(b) with an average success probability of 82.1(2)\%, indicating the Toffoli-3 is faithful for arbitrary input states.

\afterpage{\clearpage}
\newpage

\ChooseOneOracles

\ChooseTwoOracles

\end{document}